\documentclass[aps,pre,onecolumn,showpacs]{revtex4}

\usepackage{epsfig,latexsym}
\usepackage{amsmath,amscd,amssymb}

\newcommand{\be}{\begin{equation}}
\newcommand{\ee}{\end{equation}}
\newcommand{\bea}{\begin{eqnarray}}
\newcommand{\eea}{\end{eqnarray}}
\newcommand{\ran}{\rangle}
\newcommand{\lan}{\langle}
\newcommand{\bom}{\mbox{\boldmath${\omega}$}}
\newcommand{\bomsc}{\mbox{\scriptsize\boldmath${\omega}$}}

\begin{document}

\title[Coupling between denaturation and chain conformations in DNA]
{Coupling between denaturation and chain conformations in DNA: stretching,
bending, torsion and finite size effects}
\author{Manoel Manghi, John Palmeri and Nicolas Destainville}

\affiliation{Laboratoire de Physique Th\'eorique, Universit\'e de Toulouse, CNRS, 31062 Toulouse, France}
\email{manghi@irsamc.ups-tlse.fr}
\date{4 August 2008}

\begin{abstract}
We develop further a statistical model coupling denaturation and
chain conformations in DNA (Palmeri J, Manghi M and Destainville N
2007 {\it Phys. Rev. Lett.} {\bf 99} 088103). Our Discrete Helical
Wormlike Chain model takes explicitly into account the three
elastic degrees of freedom, namely stretching, bending and torsion
of the polymer. By integrating out these external variables, the
conformational entropy contributes to bubble nucleation (opening of
base-pairs), which sheds light on the DNA melting mechanism. Because
the values of monomer length, bending and torsional moduli differ
significantly in dsDNA and ssDNA, these effects are important.
Moreover, we explore in this context the role of an additional
loop entropy and analyze finite-size effects  in an experimental
context where polydA-polydT is clamped by two G-C strands, as well
as for free polymers.
\end{abstract}

\maketitle

\section{Introduction}


The study of DNA physical properties is seeing intense activity from
both a theoretical~\cite{blake,meltsim,peyrard,dauxois,
peyrardreview,hwa,peliti,carlon,nelson,blossey,jeon,marko,mazur,metzler,everaers,joyeux,JMN,JMN2} and an experimental perspective~\cite{libchaber,zocchi1,zocchi2,pouget,du,
wiggins,linna,arneodo,archer}. The first theoretical and experimental studies were published several
decades ago, but the recent development of experimental techniques
enabling one to address DNA properties at the single molecule level has
brought a significant renewal of interest in the field. They provide not only average properties like their former bulk counterparts, but also the statistics of fluctuations
around the average values. Single molecule setups range from
magnetic and optical tweezers~\cite{smith,svoboda} or Tethered Particle Motion
apparatus~\cite{finzi,pouget2,segall,nelson2}, to Atomic Force Microscopy~\cite{wiggins,ke}. They give access to huge amounts of data
concerning DNA physical properties such as bending, stretching, and
twisting elasticities or conformational dynamics~\cite{finzi,cluzel,strick,bustamante,pouget2}. In parallel, the genomic revolution leads to the elucidation of numbers of biological functions involving nucleic acids. A pressing demand follows for reliable and precise physical models, able to validate the many hypothesis emerging from molecular biology or microscopy experiments. This constitutes a double motivation for theoreticians to refine the existing microscopic DNA models: accounting for the new, accurate physics experiments; and validating (or invalidating) the physical assumptions underlying the proposed biological mechanisms.

Denaturation is one of the intimate DNA physical features that are supposed to be involved in many critical cellular functions, such as transcription, replication, protein binding, but are not fully understood. Even though DNA unwinding at the cellular level is generally an active process due to enzymes consuming energy, such as helicases~\cite{maiorano}, understanding the subtle statistical mechanics of this bio-polymer is an essential first step towards the elucidation of more complex, active mechanisms. Furthermore, the spontaneous opening of base-pairs due to thermal activation is likely to play a direct role in several biological processes. Recently, Yan and Marko~\cite{marko} have for example proposed that coupling the DNA elasticity to a minimal model of base-pair melting can account for the increased cyclization probability observed by Cloutier and Widom~\cite{cloutier}: even if it is rare, local denaturation increases short-range flexibility because single strand DNA (ssDNA) is nearly two orders of magnitudes more flexible than double strand DNA (dsDNA). This increased flexibility should play a role everywhere the polymer must be bent or looped on length scales shorter than its persistence length (typically equal to 50 nm). In the nucleosome, it is twisted around histones, the diameter of which is about 11~nm~\cite{cell_book}.

In order to get more insight into this coupling between denaturation
and elasticity, we recently proposed a more refined coupled, non-linear
model, where the internal states of base pairs (open or closed) are
described by a one-dimensional Ising model, whereas the chain
configurations are encoded by a one-dimensional Heisenberg one taking
into account DNA bending~\cite{JMN,JMN2}. By solving exactly
this model, we demonstrated that taking into account this coupling between
internal and external degrees of freedom enables the prediction of the
modifications of elastic properties when increasing the temperature: Ising
parameters are renormalized by temperature in such a way that DNA
denaturation is accompanied by a collapse of the chain persistence
length. Following this route, we were able for the first time to write the melting
temperature $T_m$ as a function of microscopic parameters only~-- when it was a fit parameter in previous models~--, and to
give a new description of boundary and finite size effects.

However, our model was minimal in that sense that only bending was
taken into account. Torsion is also known to play a role on
elasticity because a strong flexion of an elastic rod is in general
accompanied by a torsion~\cite{landau} which decreases
the energy cost of the deformation. Similarly, stretching of base
pairs ought to be included in a complete elastic model. In the present
paper, we systematically explore these effects into detail, by
proposing an exactly solvable Discrete Helical Wormlike Chain model (DHWC), and predicting how Ising parameters are renormalized in this context (Section~\ref{section2}).

In Section~\ref{section3}, we investigate the influence of the chain length (or finite-size effects) on melting profiles. At the experimental level, it has been shown in~\cite{blake,PS} that they are measurable even for DNA made of several thousand base-pairs. These effects are usually measured for polydA-polydT flanked by more stable G-C rich strands. Hence we modify our model to account for such clamped boundary conditions. In other models of denaturation~\cite{meltsim,PS,fixfr}, chain configurations are partially incorporated {\it via} a so-called ``loop entropy'' that
takes into account the entropic cost of closing a denaturation bubble
when it is not located at a polymer end. We investigate the role of loop entropy in  finite clamped and free DNA chains.


\section{Coupling between internal and external DNA's degrees of freedom}
\label{section2}


In Refs.~\cite{JMN,JMN2}, we showed that the denaturation melting temperature emerges naturally by taking into account the difference in bending rigidities of ssDNA sequences (bubbles) and dsDNA ones. Indeed, the ratio of both moduli, $\kappa_{ds}/\kappa_{ss}$ is on the order of 50. It is at the origin of an entropic barrier which stems for the fact that in the ssDNA state, the allowed spatial configurations for unit tangent vectors $\hat{\bf t}_i$, which describe the chain conformations, are much more numerous, then leading to a significant increase in entropy. More precisely, it has been shown that the free energy (mostly of entropic nature) coming out by integrating the Hamiltonian part which depends on the external variables $\hat{\bf t}_i$ renormalizes the bare Ising parameters, $K$ and $J$, which are the energy costs of creating a domain wall and destacking two adjacent base-pairs respectively. The third Ising parameter, $\mu$, which corresponds to the energy required to break a base-pair (or ``magnetic field" in a magnetism analogy), is not renormalized. In particular, the full penalty of breaking one base-pair located in DNA's interior, $L=\mu+K$, becomes
\bea
L_0 &=& \mu+K-\frac{k_BT}2\left[G_0\left(\frac{\kappa_{ds}}{k_BT}\right)-G_0\left(\frac{\kappa_{ss}}{k_BT}\right)\right]\label{bending} \\
&\simeq& \mu+K-\frac{k_BT}2\ln\left(\frac{\kappa_{ds}}{\kappa_{ss}}\right)\quad \mathrm{for}\quad \kappa\gg k_BT\nonumber
\eea
where $k_BT$ is the thermal energy and $G_0(x)=x-\ln\left(\frac{\sinh x}{x}\right)$. The approximation is valid in the temperature range of interest since $\kappa_{ss}\approx 6\,k_BT$.

In the infinitely long chain limit, the melting temperature $T_m$, defined as the temperature at which half of the base-pairs are broken, is simply given by $L_0(T_m)=0$. The melting temperature thus naturally emerges in this model and is determined by the competition between the enthalpic cost of breaking base pairs (mostly Hydrogen bonds and $\pi$-overlap of carbon ring wave-functions of adjacent nucleotides but also charge, dipolar, and Van-der-Waals interactions) and the entropic gain in nucleating bubbles made of very flexible single-stranded DNA chains.

However, other external variables than $\hat{\bf t}_i$, which also characterize the chain elasticity, may lead to a renormalization of the parameter $L$. Clearly, two other external degrees of freedom should also be taken into account:
\begin{itemize}
    \item many force-extension experiments have shown that the monomer size $a$ is no the same in dsDNA and ssDNA (see the review~\cite{bust_rev} and references therein). Indeed, the monomer size in the B-form of double-stranded DNA is generally defined as the rise along the central axis per base-pair which is $a_{ds}=0.34$~nm. The generally accepted value~\cite{marko,roland} of the monomer size in ssDNA is $a_{ss}=0.71$~nm and we choose in the following $a_{ss}\approx2\,a_{ds}$~\cite{footnote1}.
    \item the B-form of dsDNA is the famous double helix and a torsional energy has to be taken into account in a more refined model. Indeed, in the continuous Helical Wormlike chain model for DNA~\cite{yamakawa_book}, the elastic energy of the chain has two contributions: a bending term already taken into account in~\cite{JMN,JMN2} and an energy of torsional deformations which in the continuum limit reads
\be
\mathcal{E}_{\mathrm{twist}}=\frac{C}2 \int \Omega_3^2(s)\, \mathrm{d}s
\label{HWC}
\ee
where $\Omega_3={\bf \Omega}\cdot \hat{\bf e}_3$. The Darboux vector ${\bf \Omega}$ characterizes the rotation of the material frame, $\hat{\bf e}_3$ is along the molecular axis, and $s$ is the curvilinear index. The twist (or torsional) rigidity modulus $C$ has been measured in torsional experiments on dsDNA~\cite{benham2,moroz,mezard,bustamante}, and is on the order of $C_{ds}\simeq 2.4-4.5 \cdot 10^{-19}$~J.nm. The twist rigidity of ssDNA is lower because it loses its stiff helical structure and has been evaluated to be $C_{ss}\simeq9\cdot10^{-20}$~J.nm~\cite{bloomfield_book}. The ratio $C_{ds}/C_{ss}$ is on the same order of $\kappa_{ds}/\kappa_{ss}$ and will certainly modify the Ising parameters in a similar way as for the bending energy.
\end{itemize}


\subsection{Discrete Helical Wormlike Chain model}


In the present work, the DNA is modeled as a fluctuating polymer chain in a space of 3 dimensions, characterized by the \textit{external chain variables}, the set of $N$ bond vectors ${\bf t}_i$, and their orientation in space (it is thus implicitly assumed that the monomer has a three-dimensional structure); and an \textit{internal Ising variable} $\sigma_i=\pm1$ which models the internal state of dsDNA, unbroken ($U$) or broken ($B$) respectively. The modeling of the base-pair internal state by an Ising model has been developed in the 60's by Lehman, Montroll and Vedenov (see review~\cite{montroll} and references therein).

We focus on the coupling of the internal variables with the external variables which is included in the Hamiltonian part treating the fluctuating chain. A material coordinate frame is defined for each monomer $i$, $\{\hat{\bf e}_{\mu,i}\}_{\mu=1,2,3}=\{\hat{\bf u}_i,\hat{\bf n}_i, \hat{\bf t}_i\}$, where $\hat{\bf t}_i$ is the unit bond vector ${\bf t}_i={\bf R}_{i+1}-{\bf R}_i=t_i\hat{\bf t}_i$ and the two other unit vectors are in the directions of the principal axes of inertia. This triad is defined with respect to a fixed referential $\{\hat{\bf x},\hat{\bf y}, \hat{\bf z}\}$ through a rotation matrix ${\bf A}_i$ characterized by Euler angles $\bom_i=(\alpha_i,\beta_i,\gamma_i)$. The evolution of the triad along the molecular chain from monomer $i$ to monomer $i+1$ is obtained by a rotation also defined by Euler angles $(\phi_{i,i+1},\theta_{i,i+1},\psi_{i,i+1})$
\be
\hat{\bf e}_{\mu,i+1}={\bf \Lambda}_{\mu \nu}(\phi_{i,i+1},\theta_{i,i+1},\psi_{i,i+1}) \hat{\bf e}_{\nu,i}
\label{matrix_rotation}
\ee
where the rotation matrix $\bf{\Lambda}$ is the product of three rotation matrices associated with each Euler angle, but can also be viewed as the product of two rotations of angles $\theta_{i,i+1}$ and $\phi_{i,i+1}+\psi_{i,i+1}$~\cite{mezard}
\bea
{\bf \Lambda}(\phi_{i,i+1},\theta_{i,i+1},\psi_{i,i+1}) &=& R(\hat{\bf t}_i,\psi_{i,i+1})R(\hat{\bf n}_{i,i+1},\theta_{i,i+1})R(\hat{\bf t}_i,\phi_{i,i+1})\\
&=& R(\hat{\bf t}_i,\phi_{i,i+1}+\psi_{i,i+1})R(R(\hat{\bf t}_i,-\phi_{i,i+1})\hat{\bf n}_{i,i+1},\theta_{i,i+1})\nonumber
\eea
In the material coordinate frame $\{\hat{\bf e}_{\mu,i}\}$, the bond vector $\hat{\bf t}_{i+1}$ is thus defined by its spherical coordinates $(\theta_{i,i+1},\phi_{i,i+1})$. Moreover, the Euler angles $(\phi_{i,i+1},\theta_{i,i+1},\psi_{i,i+1})$ which will appear in the Hamiltonian are completely determined by the two sets of Euler angles $\bom_i$ and $\bom_{i+1}$ through ${\bf \Lambda}_{i,i+1}={\bf A}_{i+1}\cdot{\bf A}_i^{-1}$.

The configurational part of the Hamiltonian is defined as the sum of two terms
\be
\mathcal{H}[\sigma,{\bf t},\psi] =\mathcal{H}_{\rm Ising}[\sigma] + \mathcal{H}_{\rm chain}[\sigma,{\bf t},\psi]
\label{H}
\ee
where $\mathcal{H}_{\rm Ising}[\sigma]$ is the usual Ising Hamiltonian already defined in~\cite{JMN,JMN2} with three parameters $(\mu,J,K)$, and $\mathcal{H}_{\rm chain}[\sigma,{\bf t},\psi]$ is the Discrete Helical Wormlike Chain (DHWC) Hamiltonian
\bea
\mathcal{H}_{\rm Ising}[\sigma] &=& -\mu \sum_{i =1}^N \,\sigma_i - \sum_{i=1}^{N-1} \, \left[J \sigma_{i+1} \sigma_i +\frac{K}2 (\sigma_{i+1} + \sigma_i)\right]\label{ising}\\
\mathcal{H}_{\rm chain}[\sigma,{\bf t},\psi] &=&  \frac12 \sum_{i=1}^N  \frac{\epsilon_i}2 \left(|{\bf t}_i|^2-a^2_i\right)^2 + \frac12 \sum_{i=1}^{N-1}\left[\kappa_{i,i+1} (\hat{\bf t}_{i+1}-\hat{\bf t}_i)^2 \right.\nonumber \\ & & \left.+2 C_{i,i+1} (\cos\theta_{i,i+1}-\cos\lambda_{i,i+1})\right]\label{DHWC}
\eea
The first term of~(\ref{DHWC}) is a non-linear stretching term dictated by rotational and translational invariances. The values of the Lam\'e coefficient $\epsilon_i$ and the monomer length $a_i$ depend on the state of the base-pair [$(\epsilon_U,a_U)$ for $\sigma_i=+1$ and $(\epsilon_B,a_B)$ for $\sigma_i=-1$]. The second term corresponds to the bending and torsional energies. The latter can be written as $C_i \left[\mathrm{tr}{\bf \Lambda}(0,\theta_{i,i+1},0)-\mathrm{tr}{\bf \Lambda}(\phi_{i,i+1},\theta_{i,i+1},\psi_{i,i+1})\right]$, and accounts for the energy penalty associated with the twist defined by the angle $\phi_{i,i+1}+\psi_{i,i+1}$. Indeed, the angle $\lambda$ of the rotation defined in~(\ref{matrix_rotation}) is a function of $\phi+\psi$ and $\theta$ (indices ${i,i+1}$ are omitted):
\be
\cos\lambda=\frac12[\cos(\phi+\psi)(\cos\theta+1)+\cos\theta-1]
\label{gamma}
\ee
The bending $\kappa_{i,i+1}$ and torsional $C_{i,i+1}$ moduli also vary locally with the state of nearest-neighbour links [$(\kappa_U,C_U)$ for type $U-U$, $(\kappa_B,C_B)$ for $B-B$, and $(\kappa_{UB},C_{UB})$ for $U-B$]. We assume in this model that all the parameters appearing in~(\ref{H}) are independent of the nucleotide type. Hence we focus on homopolynucleotides. The case of sequence dependent parameters  could be handled numerically.

Equation~(\ref{DHWC}) defines our discrete version of the continuous Helical Wormlike Chain model first employed by Yamakawa for DNA~\cite{yamakawa_book} and extended in several articles in the literature~\cite{hearst,mezard,kamien,o'hern}. First, one observes that if there is no twist, i.e. no rotation around the tangent vector $\hat{\bf t}_i$, it imposes $\phi+\psi=0$ and from~(\ref{DHWC})--(\ref{gamma}), the torsional term vanishes. Hence if there is no twist along the chain (or if the DNA chain is modeled as linear), the DHWC becomes the classical Discrete Wormlike Chain already developed in~\cite{JMN,JMN2}. Furthermore, the Discrete Helical Wormlike Chain simplifies in the continuum limit, $x_{i+1}-x_i\to \frac{\partial x}{\partial s} \,\Delta s$ with $\Delta s\to0$ where $s$ is the curvilinear index. Indeed it is straightforward to see that $ \sum_{i=1}^{N-1}\kappa (\hat{\bf t}_{i+1}-\hat{\bf t}_i)^2\to \int \kappa [\Omega_1^2(s)+\Omega_2^2(s)]\mathrm{d}s$ and with more algebra that $ \sum_{i=1}^{N-1}C[\mathrm{tr}{\bf \Lambda}(0,\theta_{i,i+1},0)-\mathrm{tr}{\bf \Lambda}(\phi_{i,i+1},\theta_{i,i+1},\psi_{i,i+1})]$ simplifies into~(\ref{HWC}) where the Darboux vector is defined by $\hat{\bf e}_{\mu,i+1}-\hat{\bf e}_{\mu,i}\to\mathbf{\Omega}\times\hat{\bf e}_{\mu,i}$ and $\Omega_\mu(s)=\mathbf{\Omega}\cdot\hat{\bf e}_{\mu}(s)$. Finally, in the low temperature regime where the spin-wave approximation is valid ($\phi+\psi\ll1$ and $\theta\ll1$), bending and torsional contributions reduce to quadratic terms
\be
\frac12 \sum_{i=1}^{N-1}\left[ \kappa_{i,i+1}\theta_{i,i+1}^2+C_{i,i+1}(\phi_{i,i+1}+\psi_{i,i+1})^2\right]+\mathcal{O}(\theta^4,\phi^4,\psi^4)\label{spin_wave}
\ee
The discrete model defined by~(\ref{spin_wave}) has already been used in the context of DNA supercoiling~\cite{jian}.


\subsection{Stretching contribution to the entropy of  bubble nucleation}


The first stretching term in~(\ref{DHWC}) is local without any coupling between the nearest neighbours. Therefore it can be integrated out easily. The Lam\'e elastic constant $\epsilon$ is very large for DNA molecules: $\epsilon a^3$ as been evaluated as 8.4~nN for ssDNA by fitting force-extension experimental curves using ab-initio calculations~\cite{roland,footnote2}, and one can expect the same order of magnitude for dsDNA. Therefore, $\epsilon a^3\gg k_BT/a\simeq 4$~pN and the saddle point approximation applied below is valid.

By expanding the first term of~(\ref{DHWC}) and writing $|{\bf t}_i|=a_i+\delta_i$ we have
\be
(|{\bf t}_i|^2-a^2_i)^2=(|{\bf t}_i|+a_i)^2(|{\bf t}_i|-a_i)^2\approx 2 a_i^2 (|{\bf t}_i|-a_i)^2 +\mathcal{O}(\delta_i^3)
\ee
The elastic term of the Hamlitonian~(\ref{H}) simplifies into
\be
\mathcal{H}_{\rm chain}[\sigma,{\bf t},\psi] \simeq \sum_{i=1}^{N-1} \frac{\epsilon_i a_i^2}2 (|{\bf t}_i|-a_i)^2+ \kappa_{i,i+1} (1-\cos\theta_{i,i+1})+C_{i,i+1}(\cos\theta_{i,i+1}-\cos\lambda_{i,i+1})
\ee

The configurational part of the partition function is
\be
\mathcal{Z} = \sum_{\{\sigma_i\} }e^{-\beta \mathcal{H}_{\rm Ising}[\sigma]} \int \left(\prod_{i=1}^{N-1} \frac{\mathrm{d}^3{\bf t}_i \mathrm{d}\gamma_i}{8\pi^2 a_0^3} \right) \, e^{-\beta \mathcal{H}_{\rm chain}[\sigma,{\bf t},\psi]}\label{part_func1}
\ee
where $\gamma_i$ is the second twist Euler angle of ${\bf t}_i$ with respect to the reference frame and $a_0$ is a normalization length.
By using the decomposition of ${\bf t}_i$ in spherical coordinates, $(t_i,\alpha_i,\beta_i)$, one has $\mathrm{d}^3{\bf t}_i\mathrm{d}\gamma_i= t_i^2 \mathrm{d}t_i  \sin\alpha_i  \mathrm{d}\alpha_i \mathrm{d}\beta_i  \mathrm{d}\gamma_i\equiv \mathrm{d}t_i\mathrm{d}^3\bom_i$ and the partial partition function for the chain is
\be
\mathcal{Z}_{\rm chain}[\sigma] = \prod_{i=1}^N \,\int_0^\infty \frac{t_i^2 \mathrm{d}t_i}{a_0^3} \, e^{- \frac{\beta \epsilon_i a_i^2}2 (t_i-a_i)^2}  \int \prod_{i=1}^N\left(\frac{\mathrm{d}^3\bom_i}{8\pi^2}\right) \, e^{-\beta \mathcal{H}_{\rm angle}[\sigma,\bomsc]}
\ee
where $\mathcal{H}_{\rm angle}[\sigma,\bom]$ is the bending and torsional Hamiltonian.
Using the saddle point approximation for the stretching integral, we get in the large stretching constant limit
\be
\prod_{i=1}^N \,\int_0^\infty\frac{t_i^2 \mathrm{d}t_i}{a_0^3} \, e^{- \frac{\beta\epsilon_i a_i^2}2 (t_i-a_i)^2}\approx \prod_{i=1}^N \sqrt{\frac{2\pi}{\beta\epsilon_i}} \frac{a_i}{a_0^3} \equiv e^{-\sum_i^N\ln \Lambda_i}
\ee
As explained above, we assume that the stretching energy has two competitive minima for dsDNA and ssDNA. In our model it means that the elastic constant $\epsilon_i$ and the monomer size $a_i$ have two different values whether the monomer is in the unbroken ($\sigma_i=1$) or broken state ($\sigma_i=-1$).
Hence, once integrated over the local $t_i$ variables, the stretching energy part can be included in the Ising part of the Hamiltonian to get an effective Ising Hamiltonian with a renormalized $\mu$. Indeed, by defining $\ln\Lambda_i=\delta\mu\,\sigma_i+\Gamma$ where $\delta\mu=\ln\left(\frac{\Lambda_U}{\Lambda_B}\right)$ and $\Gamma=\ln(\Lambda_U\Lambda_B)$, the \textit{renormalized} temperature dependent chemical potential is
\be
\mu_0=\mu-k_BT \ln\left(\frac{a_B}{a_U}\sqrt{\frac{\epsilon_U}{\epsilon_B}}\right)
\label{mu0}
\ee
where the correction accounts for the entropic gain when the monomer state changes. It has two contributions:
\begin{enumerate}
    \item in the broken state, the monomer size is greater, $a_B\approx 2 a_U$, which implies a larger volume in the phase space and thus an increase in entropy;
    \item in the case of different elastic constants, $\epsilon_U\neq\epsilon_B$, since the stretching energy $\langle E\rangle=\frac12k_BT$ is independent of these constants, the elastic free energy difference is purely of entropic origin, similarly to the simple Einstein model for solids.
\end{enumerate}

In the present case, the elastic constants $\epsilon_U$ and $\epsilon_B$ are unknown. Although several experimental studies seem to show that the stretching constant of dsDNA is larger than for ssDNA~\cite{smith2,hegner}, we  have not been able to  find reliable values.  If, for example, we assume them equal, then the chemical potential $\mu$ is  lowered by $0.5-1\,k_BT$, which is non-negligible.


\subsection{Bending and torsional contributions}


In this section, we focus on the partition function integrated over the angles ($\mathrm{d}^3\bom_i=\sin\alpha_i  \mathrm{d}\alpha_i \mathrm{d}\beta_i  \mathrm{d}\gamma_i$). The full partition function~(\ref{part_func1}) can be written as
\be
\mathcal{Z} =  \sum_{\{\sigma_i\} } e^{-\beta \mathcal{H}_{\rm Ising,0}[\sigma]} \int \left(\prod_{i=1}^N \frac{\mathrm{d}^3\bom_i}{8\pi^2}\right) e^{-\beta \sum_{i=1}^{N-1}\kappa_{i,i+1} (1-\cos\theta_{i,i+1}) + C_{i,i+1}(\cos\theta_{i,i+1}-\cos\lambda_{i,i+1})}
\ee
where $\mathcal{H}_{\rm Ising,0}$ is the same as~(\ref{ising}) with $\mu$ replaced by $\mu_0$ given in~(\ref{mu0}). Similarly to the Discrete Wormlike Chain model~\cite{JMN,JMN2}, the partition for the coupled system can be calculated using transfer matrix techniques. For example, we have
\begin{equation}
\mathcal{Z} = \sum_{\{\sigma_i \} }  \, \prod_{i=1}^N
\,\int \frac{\mathrm{d}^3\bom_i}{8\pi^2}\langle {V|\sigma _1 } \rangle
\langle \sigma _1 |\hat P(\bom_1 ,\bom_2 )|\sigma _2\rangle
\cdots \langle \sigma _{N - 1} |\hat P(\bom_{N - 1} ,\bom_N) |\sigma _N \rangle \langle {\sigma _N |V}\rangle, \label{part_func2}
\end{equation}
where the matrix elements of the transfer kernel that appears $N-1$ times in~(\ref{part_func2}), are given by (the tilde means in units of $k_BT$)
\bea
\lan +1|\hat P(\bom_i,\bom_{i+1}) |+1\ran &=& e^{\tilde\kappa_U (\cos\theta_{i,i+1}-1)+\tilde C_U (\cos\theta_{i,i+1}-\cos\lambda_{i,i+1})+\tilde J+\tilde K+\tilde\mu_0}\\
\lan -1|\hat P(\bom_i,\bom_{i+1}) |-1\ran &=& e^{\tilde\kappa_U (\cos\theta_{i,i+1}-1)+\tilde C_U (\cos\theta_{i,i+1}-\cos\lambda_{i,i+1})+\tilde J-\tilde K-\tilde\mu_0}\\
\lan +1|\hat P(\bom_i,\bom_{i+1}) |-1\ran &=& e^{\tilde\kappa _{UB}(\cos\theta_{i,i+1}-1)+\tilde C_{UB} (\cos\theta_{i,i+1}-\cos\lambda_{i,i+1})-\tilde J}\\
&=& \lan -1|\hat P(\bom_i,\bom_{i+1}) |+1\ran
\eea
It is written in the canonical base $|U\rangle=|+1\rangle$ and
$|B\rangle= |-1\rangle$ of the U and B states. The end vector
\begin{equation}
|V\rangle=e^{\tilde\mu_0/2}|U\rangle + e^{-\tilde\mu_0/2}|B\rangle
\end{equation}
enters in order to take care of the \textit{free chain} boundary conditions~\cite{JMN2} (see also Section~\ref{section3}).

The partition function can be rewritten by examining the effective Ising model obtained by integrating over the chain conformational degrees of freedom $\bom_i$ in~(\ref{part_func2}). The problem reduces to that of an effective
Ising model with an ``effective free energy" $H_{\rm Ising, eff}$
containing renormalized parameters. This method works because, for
the coupled Ising-chain system, the rotational symmetry is not
broken. Hence the matrix obtained by integrating the kernel $\hat P(\bom_i,\bom_{i+1})$ in~(\ref{part_func2}) is the same for any site $i$.

\begin{figure}[t]
\begin{center}
\includegraphics[height=5cm]{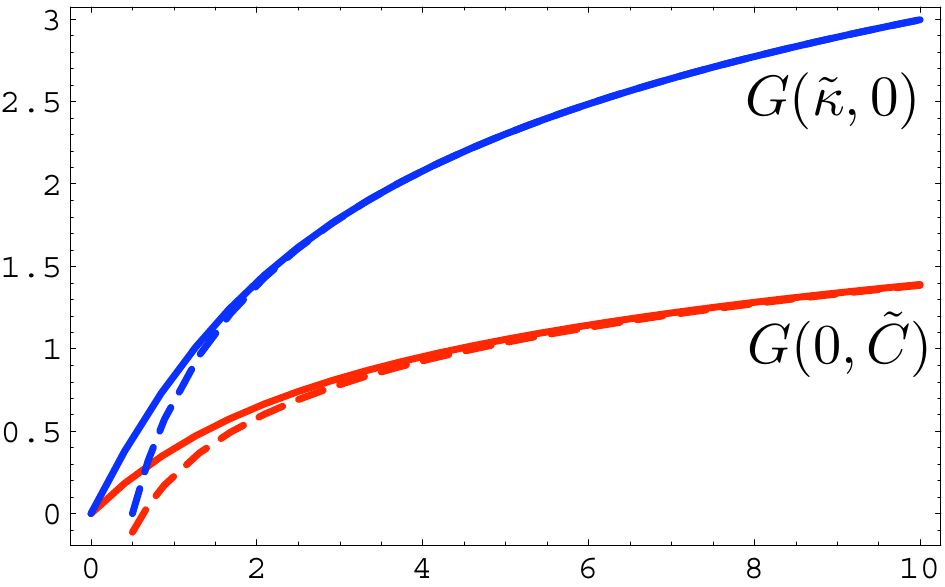}
\caption{Plots of the function $G(\tilde\kappa,0)$ and $G(0,\tilde C)$ and the asymptotic expressions~(\ref{asymptotic}) as broken lines.} \label{fig1}
\end{center}
\end{figure}

We thus are able to carry out the angle integrations in sequential fashion
by using the triad $\{\hat{\bf e}_{\mu,i-1}\}$ as the referential for
the $i^{\rm th}$ Euler angle integration. Since this corresponds for each integration to make a rotation transformation for the variables with the Jacobian equal to 1, the Euler angle  integrated transfer matrix is
\begin{equation}
\hat P_{\rm I, eff} = \int \frac{\mathrm{d}^3\bom_i}{8\pi^2}\hat
P(\bom_i,\bom_{i + 1} ) = \left( \begin{array}{*{20}c}
   {e^{-G(\tilde\kappa_U,\tilde C_U)+\tilde J+\tilde K+\tilde\mu_0}} & {e^{-G(\tilde\kappa_{UB},\tilde C_{UB})-\tilde J}}  \\
   {e^{-G(\tilde\kappa_{UB},\tilde C_{UB})-\tilde J}} & {e^{-G(\tilde\kappa_B,\tilde C_B)+\tilde J-\tilde K-\tilde\mu_0}}  \\
\end{array} \right)
\label{Pising0}
\end{equation}
where $G(\tilde\kappa,\tilde C)$ is (in units of $k_BT$) the free energy
of a single joint (two-link) subsystem with bending and torsional rigidities $(\kappa,C)$
(either $U-U$, $B-B$, $U-B$):
\bea
G(\tilde\kappa,\tilde C) &=& -\ln\left[\int \frac{\sin\theta \mathrm{d}\theta \mathrm{d}\phi \mathrm{d}\psi}{8\pi^2} e^{\tilde\kappa(\cos\theta-1)+\tilde C(\cos\theta-\cos\lambda)}\right]\label{G1}\\
&=& 2\tilde\kappa-\ln\left[\int_0^1 dx \,\mathrm{I}_0(\tilde Cx)\,e^{(2\tilde\kappa-\tilde C)x}\right], \label{G}
\eea
where $ \mathrm{I}_0$ is the modified Bessel function of the first kind~\cite{footnote3}. Two interesting cases are:
\begin{itemize}
    \item $C=0$ leading to $G(\tilde\kappa,0)=G_0(\tilde\kappa)$ already defined in~(\ref{bending}) which is an increasing function of $\tilde\kappa$ (cf. Figure~\ref{fig1}), and~(\ref{G}) is a generalization of the previous result~(\ref{bending})~\cite{JMN,JMN2};
    \item $\kappa=0$, $G(0,\tilde C)=\tilde C-\ln\left[\mathrm{I}_0(\tilde C)+\mathrm{I}_1(\tilde C)\right]$ which is also an increasing function of $\tilde C$ (cf. Figure~\ref{fig1}).
\end{itemize}
The function $G(\tilde\kappa,\tilde C)$ is plotted in Figure~\ref{fig1} which shows that it is a monotonic increasing function. In the spin-wave approximation, the integral~(\ref{G1}) is computed using the saddle-point approximation and the asymptotic behaviour of $G$ is
\be
G(\tilde\kappa,\tilde C) \underset{\tilde\kappa,\tilde C\gg1}{\longrightarrow} \ln(2\tilde\kappa)+\frac12\ln\left(2 \sqrt{\frac2\pi}\tilde C\right)
\label{asymptotic}
\ee
We observe in Figure~\ref{fig1} that the asymptotic limit is a very good approximation for $\tilde\kappa$ and $\tilde C$ larger than 2, and thus for real DNA.

The Hamiltonian of the model~(\ref{H}) then reduces to an effective Hamiltonian which is now of Ising-type
\be
\mathcal{H}_{\rm Ising,eff}[\sigma] = -\mu_0 \sum_{i =1}^N \,\sigma_i - \sum_{i=1}^{N-1} \, \left[J_0 \sigma_{i+1} \sigma_i +\frac{K_0}2 (\sigma_{i+1} + \sigma_i)\right]
\label{ising_effective}
\ee
where the bare Ising parameters $K$ and $J$ are renormalized according to
\bea
K_0 &=& K-\frac{k_BT}2[G(\tilde\kappa_U,\tilde C_U)-G(\tilde\kappa_B,\tilde C_B)]\label{K0}\\
J_0 &=& J-\frac{k_BT}4[G(\tilde\kappa_U,\tilde C_U)+G(\tilde\kappa_B,\tilde C_B)-2G(\tilde\kappa_{UB},\tilde C_{UB})]\label{J0}
\eea
and $\mu_0$ is defined in~(\ref{mu0}).

Usually, it is admitted that the torsional modulus is proportional to the bending modulus $C\simeq1.6\,\kappa$~\cite{mezard}. Taking the same values as in~\cite{JMN,JMN2} for a polydA-polydT homopolymer, $\tilde\kappa_U=\tilde\kappa_{UB}=147$ and $\tilde\kappa_B=5.54$ at $T=T_m=326$~K, we get $G(\tilde\kappa_U,\tilde C_U)=9.3$ and $G(\tilde\kappa_B,\tilde C_B)=4.3$ which leads to a decrease of $K$ and $J$ by about $2-3\,k_BT$ and $1-2\,k_BT$ respectively in the temperature range of interest. We have found in~\cite{JMN2} $\mu=1.78 \,k_BT$, $J=3.64 \,k_BT$ and $K$ was set to 0. Hence these entropic contributions are on the same order of magnitude as the bare values and must be taken into account.

Moreover, with these values, the spin-wave approximation applies and we can summarize~(\ref{mu0}) and~(\ref{K0}) as
\be
L_0 =\mu_0+K_0 \approx \mu+K-\frac{k_BT}2\ln\left(\frac{a_B^2\epsilon_U\kappa_U\sqrt{C_U}}{a_U^2\epsilon_B\kappa_B\sqrt{C_B}} \right)
\ee
showing that the renormalization of the Ising parameters comes essentially from entropic effects, namely stretching, bending and torsional entropies.

This twist-induced melting might be important in the context of single molecule torque experiments~\cite{smith,cluzel,strick,bustamante,cocco} and in the context of superhelical stressed circular dsDNA~\cite{benham}. For instance, within this model, applying a torque (or a twist) will locally modify the free energy cost $L$ to nucleate a bubble and will, in return, influence the mechanical response of the chain.

In the rest of the paper, we will be interested in expectation values depending only on the spin variables $\sigma_i$. Hence, everything can be computed using directly the effective ising Hamiltonian~(\ref{ising_effective}) with renormalized parameters, $\mu_0$, $L_0$ and $J_0$. In principle, the DHWC model could be completely solved by transfer matrix techniques, thus requiring the diagonalization of the transfer operator $\hat P(\bom_i,\bom_{i+1})$ defined in~(\ref{part_func2}). This out of the scope of the present work.


\subsection{End-to-end distance}


In this section, we compute the end-to-end distance of a dsDNA using the model presented in~\cite{JMN} where we neglect the torsional term. We show that the difference in monomer sizes in the unbroken and broken states modifies the end-to-end distance and should be taken into account. Therefore, we complete the findings of~\cite{JMN2} where the monomer sizes were supposed to be equal.

The end-to-end distance of the chain is defined as $R=\sqrt{{\bf R}^2}$, where
\bea
{\bf R}^2 &=& \sum_{i,j=1}^N \langle (a_i \hat{\bf t}_i) \cdot (a_j \hat{\bf t}_j)\rangle \\
&=& \sum_{i,j=1}^N A^2\langle \sigma_i {\bf t}_i \cdot {\bf t}_j \sigma_j\rangle + AB \left( \langle \sigma_i {\bf t}_i \cdot {\bf t}_j\rangle +\langle {\bf t}_i \cdot {\bf t}_j \sigma_j \rangle \right) + B^2 \langle {\bf t}_i \cdot {\bf t}_j\rangle\nonumber
\eea

\begin{figure}[ht]
\begin{center}
\includegraphics[height=6cm]{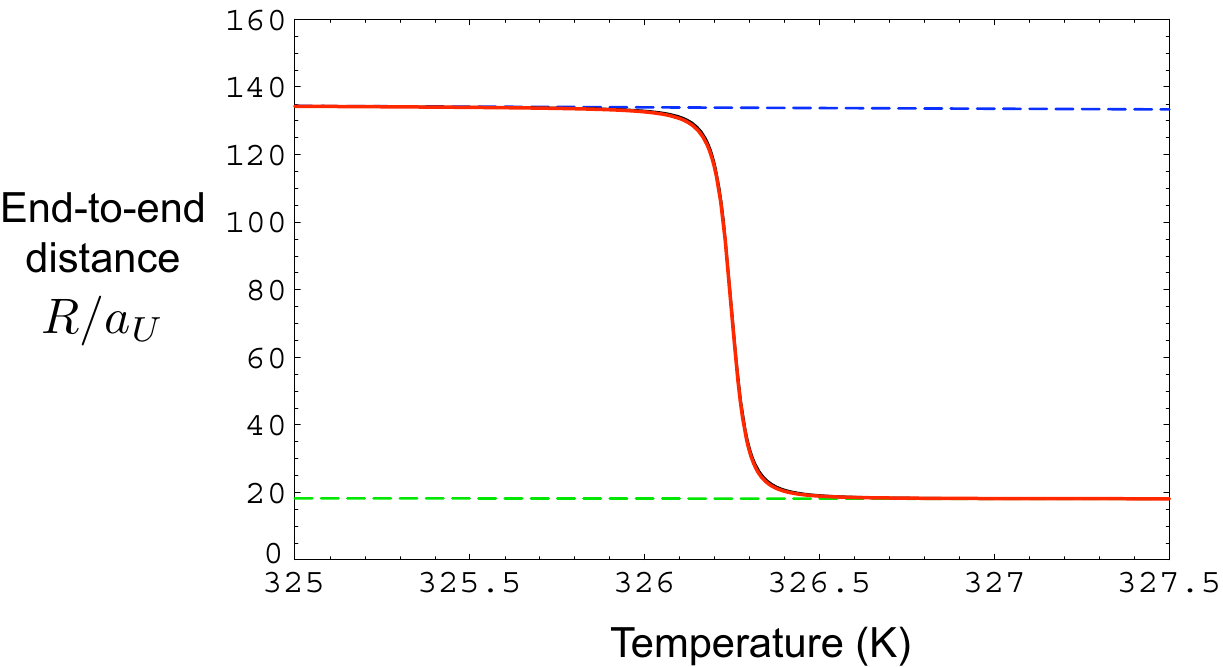}
\caption{End-to-end distance (in units of base-pair size) as a function of the temperature $T$ for the parameter values $\mu=4.46$~kJ/mol, $J=9.13$~kJ/mol, $K=0$, corresponding to $T_m=326.4$~K and $\tilde\kappa_U=\tilde\kappa_{UB}=147$, $\tilde\kappa_B=5.5$, $a_B=2a_U$. The full calculation~(\ref{endtoend}) (in red) and the interpolation formula~(\ref{interpol}) (in black) coincide. The blue and green broken lines correspond to the bare dsDNA and ssDNA respectively.} \label{fig2}
\end{center}
\end{figure}

The monomer size, which depends on  the internal variable $\sigma_i$, can be written as $a_i=A\sigma_i+B$ with $A=(a_U-a_B)/2$ and $B=(a_U+a_B)/2$. In the thermodynamic limit, $N\to\infty$, this expression simplifies to
\bea
\frac{{\bf R}^2}{N} &\underset{N\to\infty}{\longrightarrow}& \left(A^2\langle\sigma_i^2\rangle +2AB \langle\sigma_i\rangle +B^2\right) + 2 \sum_{r=1}^\infty \left[ A^2\langle \sigma_i {\bf t}_i \cdot {\bf t}_{i+r} \sigma_{i+r}\rangle\right.\\ & & \left. + AB \left(\langle \sigma_i {\bf t}_i \cdot {\bf t}_{i+r} \rangle +  \langle {\bf t}_i \cdot {\bf t}_{i+r} \sigma_{i+r}\rangle\right) + B^2  \langle {\bf t}_i \cdot {\bf t}_{i+r}\rangle \right]\nonumber
\eea
which is independent of $i$. By using the transfer matrix approach and the results already presented in~\cite{JMN2}, we find after some lengthy calculations
\bea
\frac{{\bf R}^2}{N} &\underset{N\to\infty}{\longrightarrow}& A^2 +2AB \lan\hat\sigma_z\ran +2 B^2 \xi^p_{\rm eff}\label{endtoend}\\
& & +2 \sum_\tau \left(A^2 \lan 1,\tau|\hat \sigma_z|0,+\ran^2+ 2 AB \lan 0,+|1,\tau\ran\lan 1,\tau|\hat\sigma_z|0,+\ran \right)\frac{e^{-1/\xi_\tau}}{1-e^{-1/\xi_\tau}}\nonumber
\eea
where the effective persistence length is defined as
\be
\xi^p_{\rm eff} \equiv \frac12 \sum_\tau \lan 1,\tau | 0, + \ran^2 \frac{1 + e^{-1/\xi_\tau}}{1-e^{-1/\xi_\tau}}
\ee
The Pauli matrix $\hat \sigma_z$ acts only on the second part of the basis that diagonalizes the transfer matrix operator $\hat P$: $|\Psi_{l,m,\tau}\ran= |l,m\ran\otimes|l,\tau\ran$ (where $(l,m)$ are the quantum numbers associated to the spherical harmonics and $\tau=\pm$ labels the eigenstates of the Ising model). In the basis $|0,\pm\ran$ we have
\be
\hat \sigma _z= \left( \begin{array}{*{20}c}
   \lan c\ran_\infty & \sqrt{1-\lan c\ran_\infty}  \\
   \sqrt{1-\lan c\ran_\infty} & {-\lan c\ran_\infty}  \\
\end{array} \right).\label{sigmaz}
\ee
where $\lan c\ran_\infty$ is the expectation value of average spin variable (or ``magnetization") in the thermodynamic limit
\be
\frac1N \sum_1^N \lan\sigma_i\ran \underset{N\to\infty}{\longrightarrow} \left\langle c \right\rangle_\infty = \frac{\sinh(L_0)}{[\sinh^2(L_0) + e^{-4J_0}]^{1/2}} \label{c}
\ee
The parameter $L_0$ is defined in~(\ref{bending}) and $J_0$ in~(\ref{J0}) setting $\tilde C=0$ for the three cases. The two orthonormal eigenvectors for a fixed $l$ are defined in~\cite{JMN,JMN2}.

The result~(\ref{endtoend}) is shown in figure~\ref{fig2} for $a_B=2a_U$. An accurate interpolating formula is given by
\be
{\bf R}_{\rm interpol}^2 = 2 N (\varphi_U a_U^2  \xi^p_U + \varphi_B a_B^2 \xi^p_B) = (1-\varphi_B) {\bf R}_{\rm ds}^2 + \varphi_B{\bf R}_{\rm ss}^2
\label{interpol}
\ee
thus generalizing a similar result given in~\cite{JMN2} for the case $a_U=a_B$.


\section{Finite size effects within the DHWC model}
\label{section3}


In this section, we study the behaviour of the fraction of open
base-pairs, $\varphi_B (N, T)$, as a function of both temperature
and chain length for  homogeneous DNA with free and modified
boundary conditions (necessary for DNA inserts). Despite early
recognition \cite{PS1} that a careful experimental study of such
homogeneous DNA polymers of varying length would be of great help in
advancing our theoretical understanding of DNA denaturation,
unfortunately such a study has not yet been carried out. As a
consequence, important questions concerning the competition between
end unwinding and internal bubble formation for finite chains, as
well as the correct form of the loop entropy factor (including the
effect of chain rigidity) and the role of chain disassociation,
remain open. Our goal here is to shed further light on the role of
polymer length in the thermal denaturation homogeneous DNA (see
\cite{sahin} for recent study of finite size effects within the
framework of generalizations to the Peyard-Bishop model).

The model we use here is a generalization of the one presented in
\cite{JMN, JMN2} and has been defined in Section~\ref{section2}. The
renormalized chemical potential is given by (\ref{mu0}) and for purposes
of illustration we use the simpler, but accurate, spin-wave
approximations for the two other renormalized parameters, summarized
here:
\begin{eqnarray}
L_0 (T)&  \approx &
L-\frac{k_BT}2\ln\left(\frac{a_B^2\epsilon_U\kappa_U\sqrt{C_U}}{a_U^2\epsilon_B\kappa_B\sqrt{C_B}}
\right) \label{L0}\\
J_0 (T) & \approx & J -\frac{k_BT}{4}\ln\left(\frac{\kappa_U
\kappa_B}{\kappa_{UB}^{2}}
    \frac{ \sqrt{ C_U C_B }}{C_{UB} }
\right),
\end{eqnarray}
where $L=\mu + K$. We also adopt the following physically reasonable
set of  model parameters: $\kappa_U/\kappa_{B}=147/5.5=26.7$,
$\kappa_{UB}=\kappa_{U}$, $a_B/a_U=\epsilon_U/\epsilon_B=2$, and
$C_B/\kappa_{B}=C_U/\kappa_{U}=C_{UB}/\kappa_{UB}=1.6$. When loop
entropy is not included in the model we use a value for $J$ obtained
previously by fitting experimental melting data for a
homopolynucleotide polydA-polydT : $J=9.13$~kJ/mol~\cite{JMN, JMN2}
(we recall that the renormalized value $J_0$ is a key parameter in
determining the transition width). When the effect of loop entropy
on the thermal denaturation of free chains is studied, we will use a
smaller value of $J$, half of the larger one, as it is well known
that loop entropy tends to sharpen the transition~\cite{PS,
montroll, PS1}. If the model prediction (without loop entropy) for the
melting temperature for polydA-polydT of length $N=30000$ base-pairs
is chosen to agree with the experimental results in~\cite{blake}
($T_m^{\rm expt.} = 338.70$~K), then we obtain $L = \mu + K =
9.87$~kJ/mol,  close to the value obtained by setting $L_0
(T_m^{\rm expt.}) = 0$ (which gives the model result without loop
entropy for the infinite chain melting temperature, see
Eq.~\ref{L0}). Using $J=9.13$~kJ/mol, we find that $J_0 \simeq
12.3$~kJ/mol at $T=339$~K, which implies that the entropic
contribution is greater than 25~\% near the melting temperature [$\tilde{J}_0 = \beta J =
3.23$ for  $\beta = 1/(k_B T_m^{\rm expt.})$].
Using $J=4.57$~kJ/mol, we find that $J_0 \simeq 7.70$~kJ/mol at
$T=339$~K, which gives an entropic contribution of 41~\%.

In our previous work \cite{JMN,JMN2} we assumed that the difference in bare stacking
energy, $K$, between the U and B states was zero. This choice was
based on evidence that near room temperature single stranded polyrA
remains stacked~\cite{krueger}. It seems, however, that near the dsDNA melting
temperature dT single strands are probably completely  and
dA ones partially unstacked~\cite{blake} with an unstacking fraction
close to 75\% near $T_m$~\cite{blake}. We can conclude that the
single dT and dA strands in polydA-polydT bubbles may have much less
stacking energy than the helical segments and  incorporate this
effect into the model by introducing a weighting parameter, $f$,
that measures the contribution of $K$ to $L$ at fixed $L$:  $K = fL$
and $\mu = (1-f)L$. Although the two unbound single dT and dA
strands in a polydA-polydT bubble may not behave exactly like two
free single dT and dA strands, the above discussion does suggest
that $f$ may be large near the melting temperature. Indeed, if we
accept the putative experimental value for the bare enthalpy needed
to open one A-T base-pair as a measure of $\mu$, then we find
$\mu\simeq 5.25$~kJ/mol~\cite{JMN,JMN2,pincet}. Using this result and
the above value for $L$ then yields $f\simeq0.5$. When $f$ is taken
to be zero there is no loss in stacking energy when a bubble opens
and we recover the case previously studied in~\cite{JMN, JMN2}.

An important question is how to incorporate bubble loop entropy into
statistical models of fluctuating DNA. This loop contribution arises
from the extra cost in free energy (with respect to two single
unbound end chains) needed to form a closed loop of bases making up
a bubble~\cite{PS, PS1, gotoh, wartell}. When loop entropy is neglected
Poland-Scheraga (PS) type models reduce to effective Ising ones,
albeit without the end-interior asymmetry that naturally arises
within our approach from the difference between $L_0$ and $\mu_0$
(see Eq.~20 of \cite{JMN2}). This can arise both from a
dissimilarity between $\mu$ and $K$ and from the renormalizations
coming from integrating out the conformational degrees of freedom.
If  without justification we formally set $\mu_0$ equal to $L_0$ we
recover previous Ising/PS type models without loop entropy.

For finite DNA polymers, end effects may have a strong influence on
both the thermal denaturation transition and chain conformational
properties.  As already discussed in~\cite{JMN2} the coupled DNA
model that we have developed is extremely useful for investigating
the dependence of various system properties on chain length, $N$.
For DNA homopolymers two types of situations can be envisaged: (i)
finite homopolymers with free end boundary conditions, and (ii)
finite polydA-polydT inserts between more stable G-C rich domains
with much higher melting temperatures.

Case (i) has already been extensively studied theoretically
in~\cite{JMN2} when the loop entropy associated with bubbles is
neglected. Although for very long chains end effects are unimportant
and  $f$ plays no role (only the value of $L$ is important), for not
too long finite chains $f$ has a strong influence on the melting
curves. Within the scope of our model with $f=0$ it was found
previously that for finite DNA chains thermal denaturation takes
place in an inhomogeneous fashion with the probability of base-pair
opening being higher at chain ends for temperatures $T < T^*$. At
the temperature $T^*$  the fraction of broken base-pairs becomes
independent of chain length and the probability of base-pair opening
becomes independent of position on the chain (see Figs.~6 and 7
of~\cite{JMN2}). For $f=0$ it was also found that the melting
temperature obeys $T^* < T_m^\infty < T_m (N)$ [where
$T_m^\infty=T_m(N\to\infty)$] and, along with the transition width,
decreases with increasing $N$. For $T < T^*$ the fraction of open
base-pairs, $\varphi_B (N)$, decreases  with increasing $N$, whereas
for $T > T^*$, it increases with increasing $N$. We further this
previous theoretical study here by investigating the influence of
the weighting factor $f$.

Without loop entropy previous Ising/PS type models predict $T_m$
independent of $N$ and therefore $T_m = T^*$. When loop entropy is
added to these models $T_m(N)$ becomes an increasing function of
$N$, which appears to agree with experiment (in \cite{blake} it was
found that the melting transition for a free homopolymer of length
$N=30000$ takes place at a temperature 1~K higher than that of
chains of length $N \approx 500$ and is much sharper). We also
examine in detail the validity of the one-sequence approximation for
free boundary conditions and investigate the influence of loop
entropy in situations where the accuracy of this approximation can
be gauged~\cite{PS, PS1}. For free boundary conditions this approximation
involves keeping only the base-pair states forming one interior
bubble or one helix section of variable length $0 \leq n \leq N$.
Unfortunately, there do not appear to be any detailed experimental
studies of the thermal denaturation of DNA homopolymers with free
ends as a function of chain length (see, however~\cite{montroll, wartell1})
that can be used to test the model predictions and clarify the role
and importance of both end effects and bubble loop entropy.

For the case (ii) of an  A-T insert of length $N$ in larger more
stable DNA polymers, detailed experiments~\cite{blake} have already
been carried out for $60 < N < 140$  and also interpreted using both
a simple two-state approximation for the A-T insert and the
Poland-Scheraga model~\cite{PS} (including loop entropy) for the
entire polymer~\cite{blake}. For inserts the boundary conditions are
fixed mainly by the exterior G-C rich domains and only $L$ enters
(and not $f$, i.e., the individual values of $\mu$ and $K$). For
inserts the one-sequence approximation involves keeping only the
base-pair states forming one bubble of variable length $0 \leq n
\leq N$. The two-state approximation accounts only for the
completely closed and the completely open chain states in the
partition function~\cite{PS} and is a special case of the more
general one-sequence approximation. The validity of these types of
approximations relies intimately on the relatively large cost in
free energy for creating a bubble (or base-pair domain walls)
compared with the cost of changing the length of an already existing
bubble (i.e., $|L_0| \ll J_0$). The upshot is that a one-bubble
state can have a variable length (and in dynamics undergoes
breathing) and such states should dominate the free energy for not
too long chains (and for longer chains, temperatures not too close
to the melting one).

We reexamine this problem by analyzing the same experimental
results~\cite{blake} using our coupled model for a finite chain with
modified boundary conditions, because in such situations the nature
of end monomers becomes extremely important. In doing so, we study
the validity of both the two-state and one-sequence approximations
without loop entropy by comparing the predictions of these
simplified approaches to those obtained from the exact solution to
our model. By incorporating the loop entropy into the one-sequence
approximation, we also examine the role and importance of this
effect for homopolymer inserts. In order to compare the predictions
of the model with experiments on A-T inserts we have fitted the
DNA melting data presented in Fig.~6 and 7 of~\cite{blake}
using simple fitting functions, the goal being to get a smooth
approximation to the data (see Appendix) that will be useful in this section.

\subsection{Exact results for General Chain Boundary Conditions (without loop entropy)}

Using transfer matrix techniques we have shown that it is possible
to obtain a compact expression for the average fraction of open
base-pairs in a finite chain of length $N$ for arbitrary boundary
conditions~\cite{JMN2} (with neither loop entropy, nor chain
sliding):
\begin{equation} \label{phin}
\varphi_B (N, T; \tilde\mu') = \frac{1}{2} \left[ 1 - \langle
c\rangle (N,T; \tilde\mu') \right]
\end{equation}
where $\langle c\rangle (N, T; \tilde\mu') \equiv 1/N \sum_{i =
1}^N\,\langle \sigma_i \rangle$ is given by
\begin{equation} \label{cn}
\langle c\rangle (N, T; \tilde\mu') = \langle c\rangle_\infty \left[
1 - \frac{2 R_V^{2}}{R_V^{2}+ e^{(N-1)/\xi_I} } \right] + \frac{2
R_V \sqrt{1-\langle c\rangle_\infty^2} \left( 1- e^{-N/\xi_I}
\right)}{N \left[ 1+R_V^2 e^{-(N-1)/\xi_I}  \right]
\left(1-e^{-1/\xi_I}\right)}
\end{equation}
$\xi_{I}$ is the Ising correlation length, and
\begin{equation}
R_V (\tilde\mu') \equiv \frac{\langle V' |0,- \rangle }{ \langle V'
|0,+\rangle} \label{RV}
\end{equation}
with the normalized end vector
\begin{equation}
|V' (\tilde\mu') \rangle= \left[ 2 \cosh(\tilde\mu') \right]^{-1/2}
\left(e^{\tilde\mu'/2}|U\rangle + e^{-\tilde\mu'/2}|B\rangle \right)
\label{endv}
\end{equation}
enforcing the chain boundary conditions. The quantities $\langle
c\rangle (N, T; \tilde\mu')$, $\langle c\rangle_\infty$ given in~(\ref{c}), $R_V
(\tilde\mu')$, and $\xi_{I}$ are all functions of $L_0$
and $J_0$~\cite{JMN2}. For free ends $\tilde\mu' = \tilde\mu_0$,
whereas for closed (open) ends, $|V'\rangle=|U\rangle$
($|B\rangle$), which can be seen by taking the $\tilde\mu'
\rightarrow \pm \infty$ limits of (\ref{endv}). When $\tilde\mu'$ is formally set equal to
$\tilde L_0$ there is no longer any end-interior asymmetry and the
model reduces to older Ising/PS type \cite{montroll} models without loop entropy.

\begin{figure}[ht]
\begin{center}
\includegraphics[height=4cm]{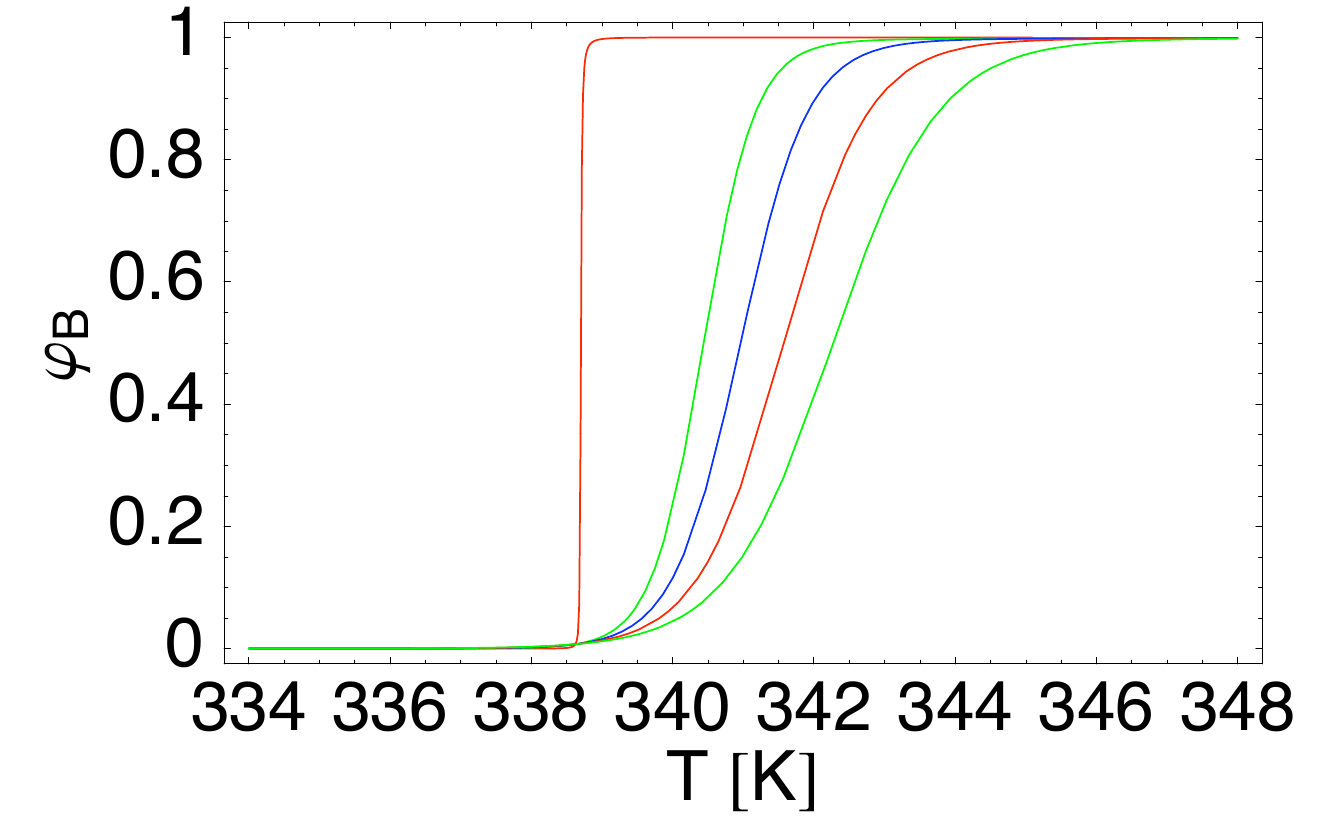}(a)
\includegraphics[height=4cm]{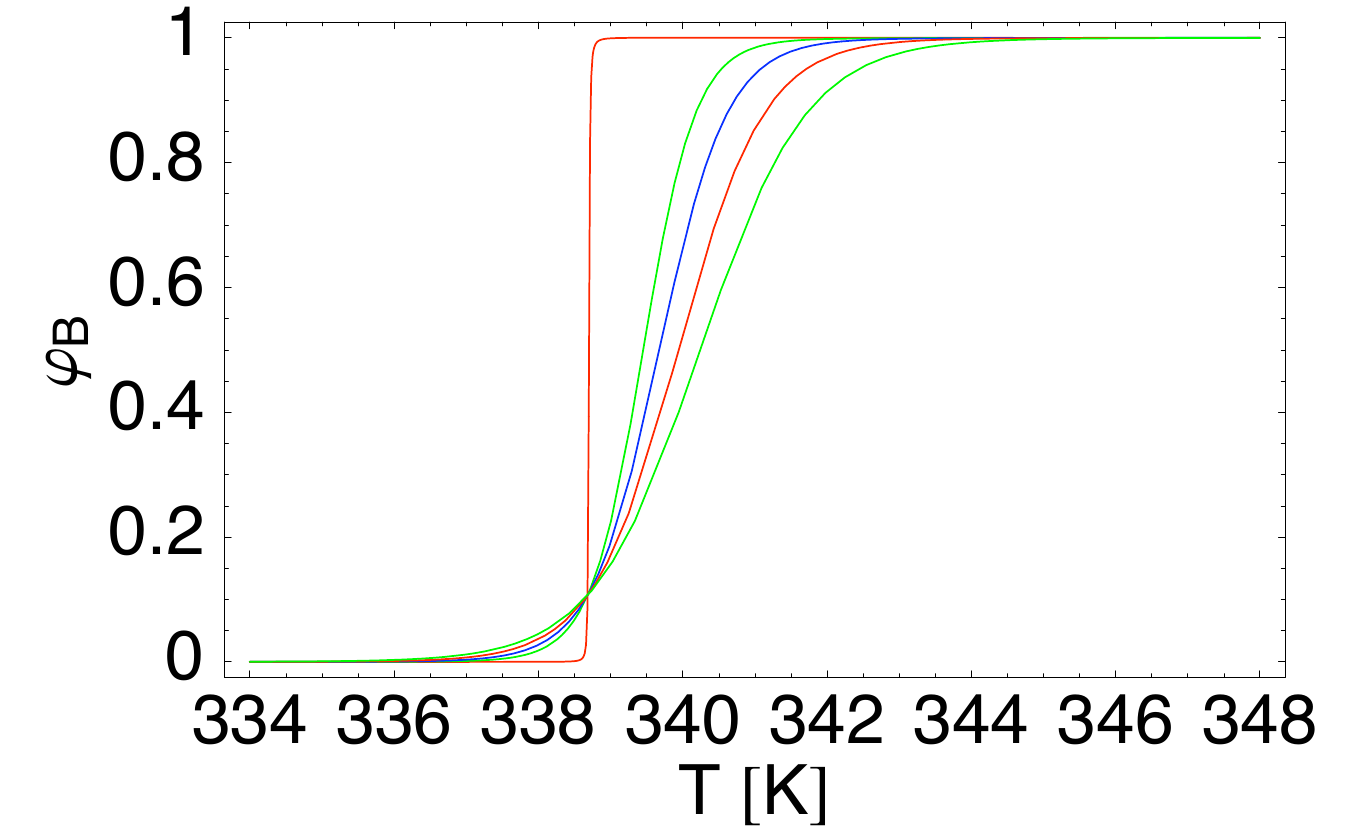}(b)
\includegraphics[height=4cm]{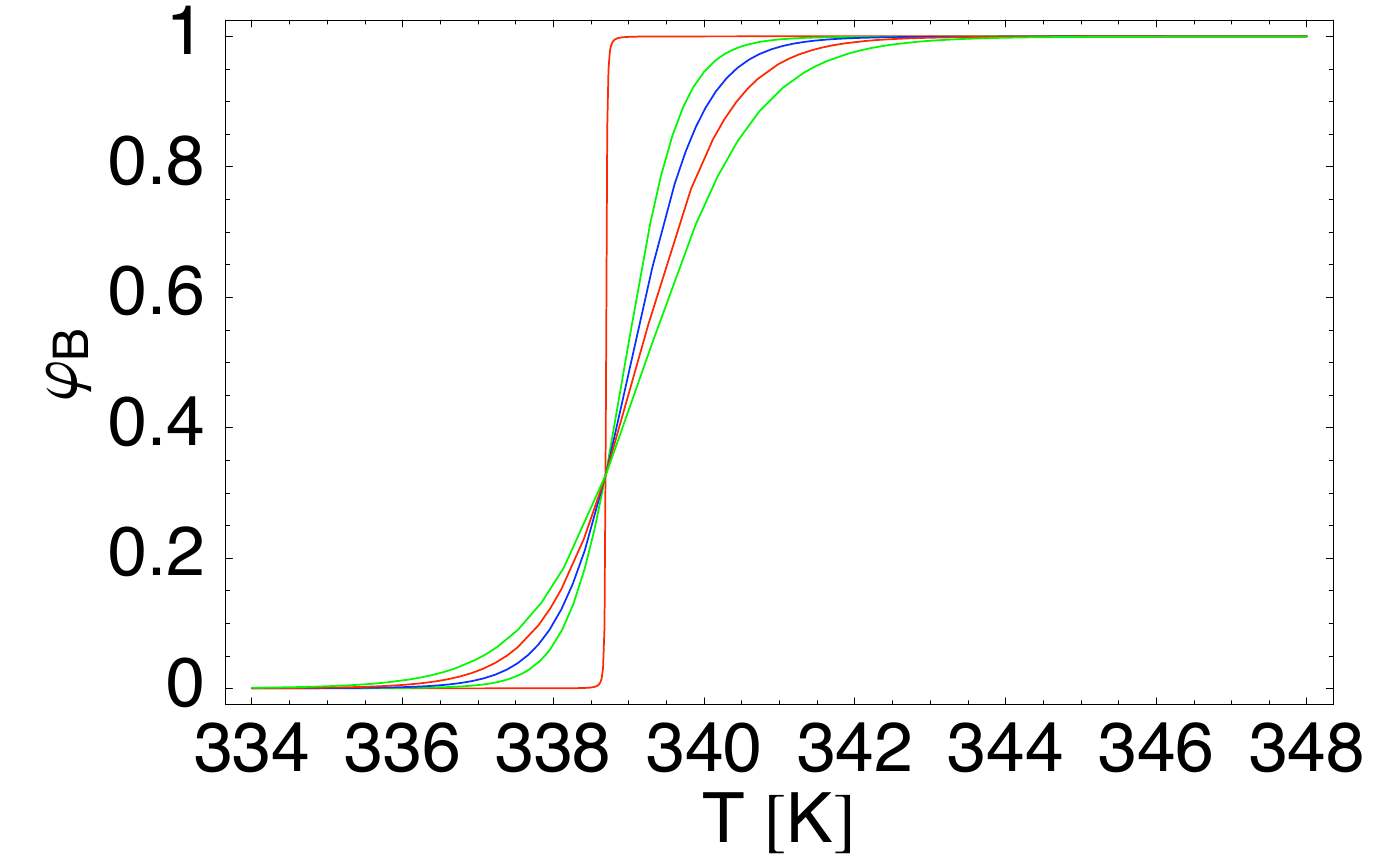}(c)
\includegraphics[height=4cm]{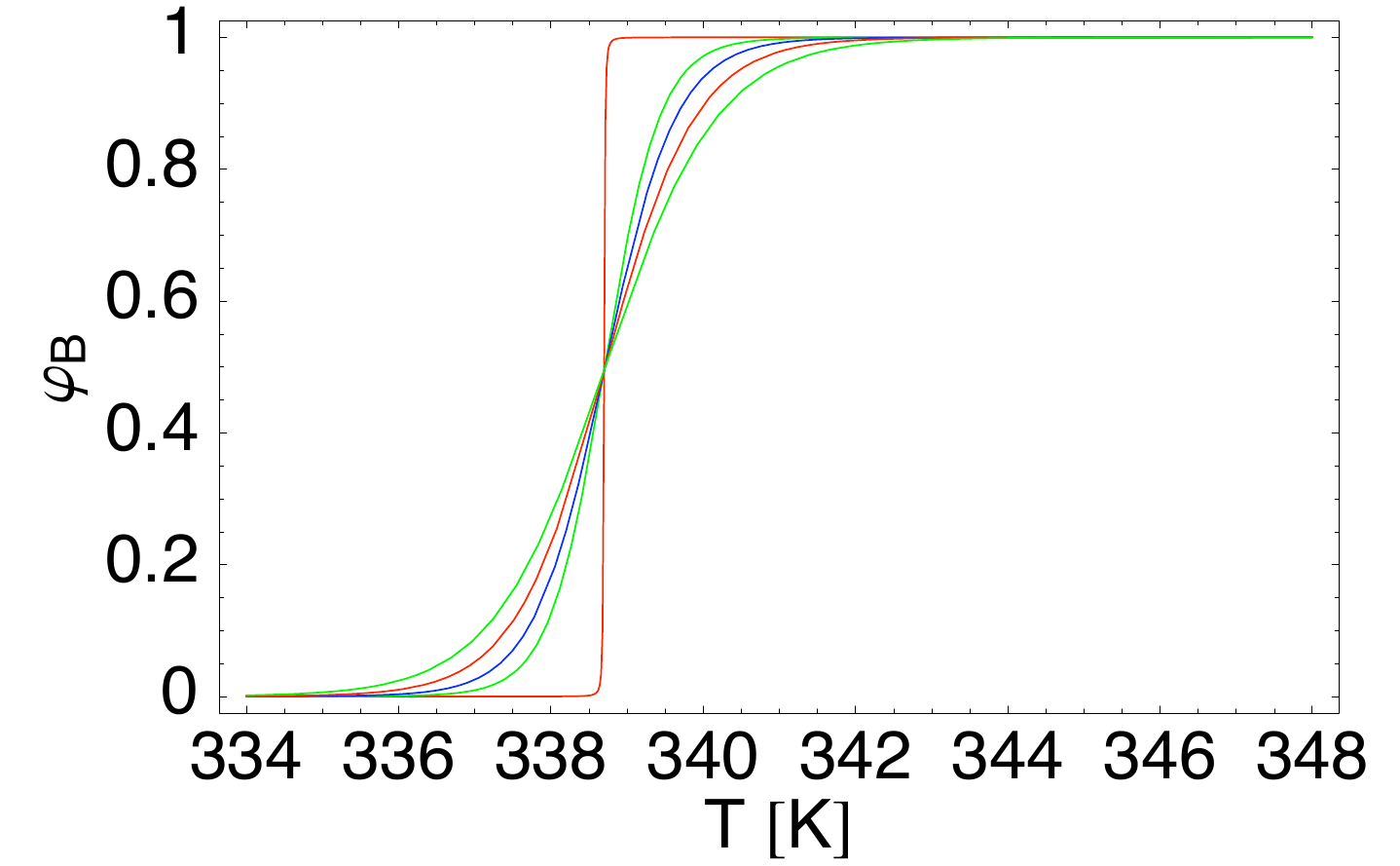}(d)
\includegraphics[height=4cm]{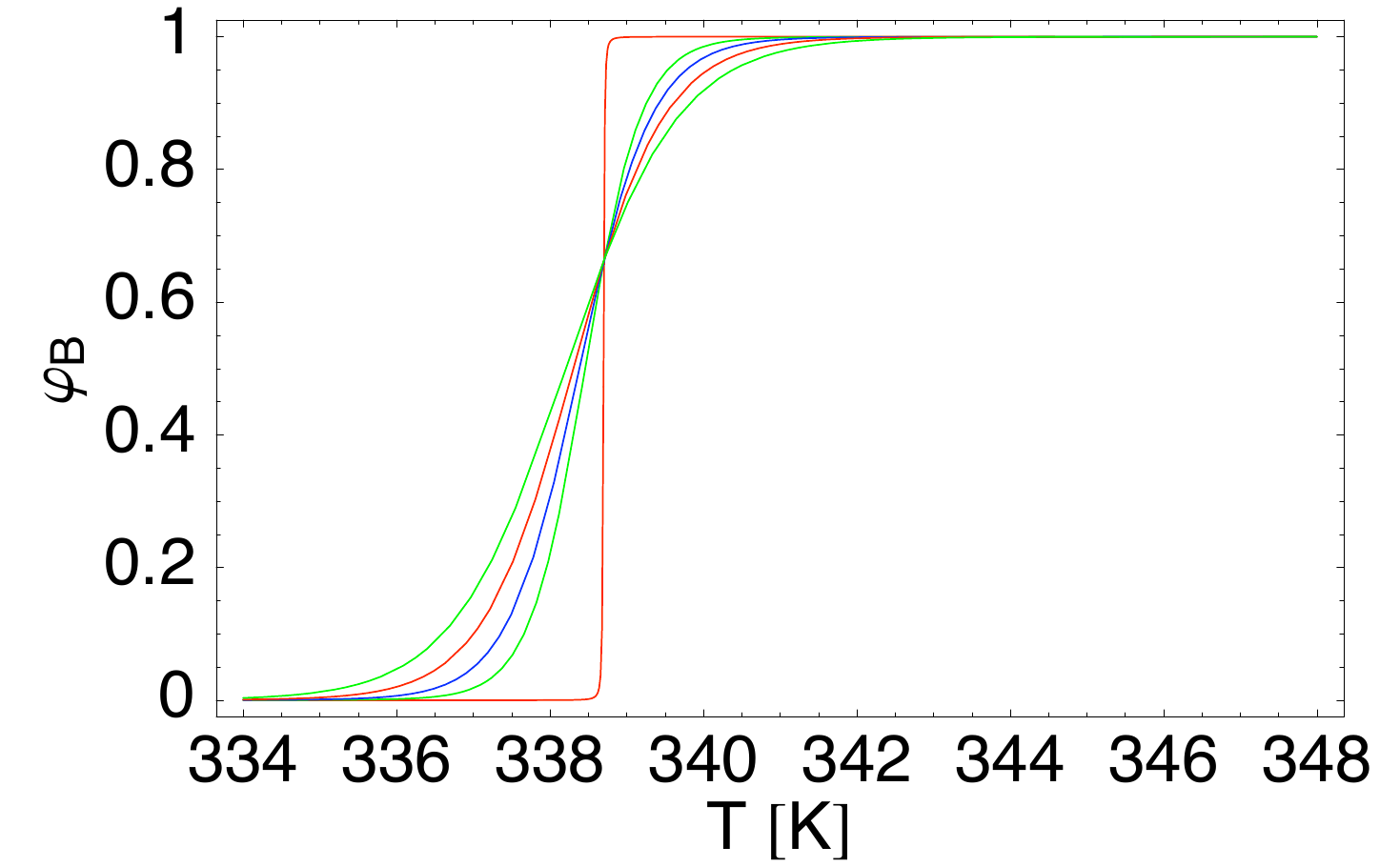}(e)
 \caption{Fraction of broken base-pairs~(\ref{phin}) vs. temperature for free boundary conditions (without loop entropy) and
chain lengths of  $N=30000$, 136, 105, 83, and 67 (from left to
right, above the temperature of intersection, $T^*$). (a) $f=0$, (b)
0.4, (c) 0.6, (d) 0.7, (e) 0.8 (other model parameters used are listed at the beginning of Section~\ref{section3}).} \label{fig3}
\end{center}
\end{figure}

A simple expression can be obtained for $R_V$ by setting $N=1$
in~(\ref{cn}) and solving for $R_V$:
\begin{equation}
R_V (\tilde\mu')  = \frac{\langle c\rangle_1- \langle
c\rangle_\infty}{ \sqrt{1-\langle c\rangle_\infty^2}  +
\sqrt{1-\langle c\rangle_1^2}}
\end{equation}
where  $\langle c\rangle_1 = \tanh(\tilde\mu')$ is a function of
$\tilde\mu'$ and therefore reflects  the boundary conditions.

\subsubsection{DNA Chains with free boundary conditions}

When $\tilde\mu' = \tilde\mu_0$ (free boundary conditions),
$R_{V,\rm{free}} = R_{V} (\tilde\mu_0)$ gets simplified in the
following way for special values of $T$~\cite{JMN2}:
\begin{equation}\label{specialv}
R_{V,\rm{free}} =  \left\{
\begin{array}{ll}
    -e^{-\tilde\mu_0},  &  T <  T^*  \\
   0,   &  T =  T^*  \\
   \tanh(\tilde\mu_0/2),  &  T =  T_m^\infty  \\
    e^{\tilde\mu_0},  &  T >  T_m^\infty \\
\end{array} \right.
\end{equation}
which shows that $R_{V,\rm{free}}$ is a monotonically increasing
function of $T$ and vanishes at  $T=T^*$.

In Fig.~\ref{fig3} we present model  results (with neither loop entropy, nor
chain sliding) based on Eq.~(\ref{cn})  for free chains of different
lengths and different values of $f$. We observe that $T^*$ increases with increasing $f$; for  $f<0.7$,
$T_m(N)$ decreases with increasing $N$, whereas for $f>0.7$,  $T_m(N)$
increases with increasing $N$. When $f \approx0.7$, the melting
curves are nearly identical with the results obtained from older
Ising/PS type models ($\mu' = L_0$) without loop entropy. When
loop entropy is added to the model the melting temperatures for the
longer chains will be shifted to the right amplifying the effect of
finite $f$ (see below).

\begin{figure}[ht]
\begin{center}
\includegraphics[height=6cm]{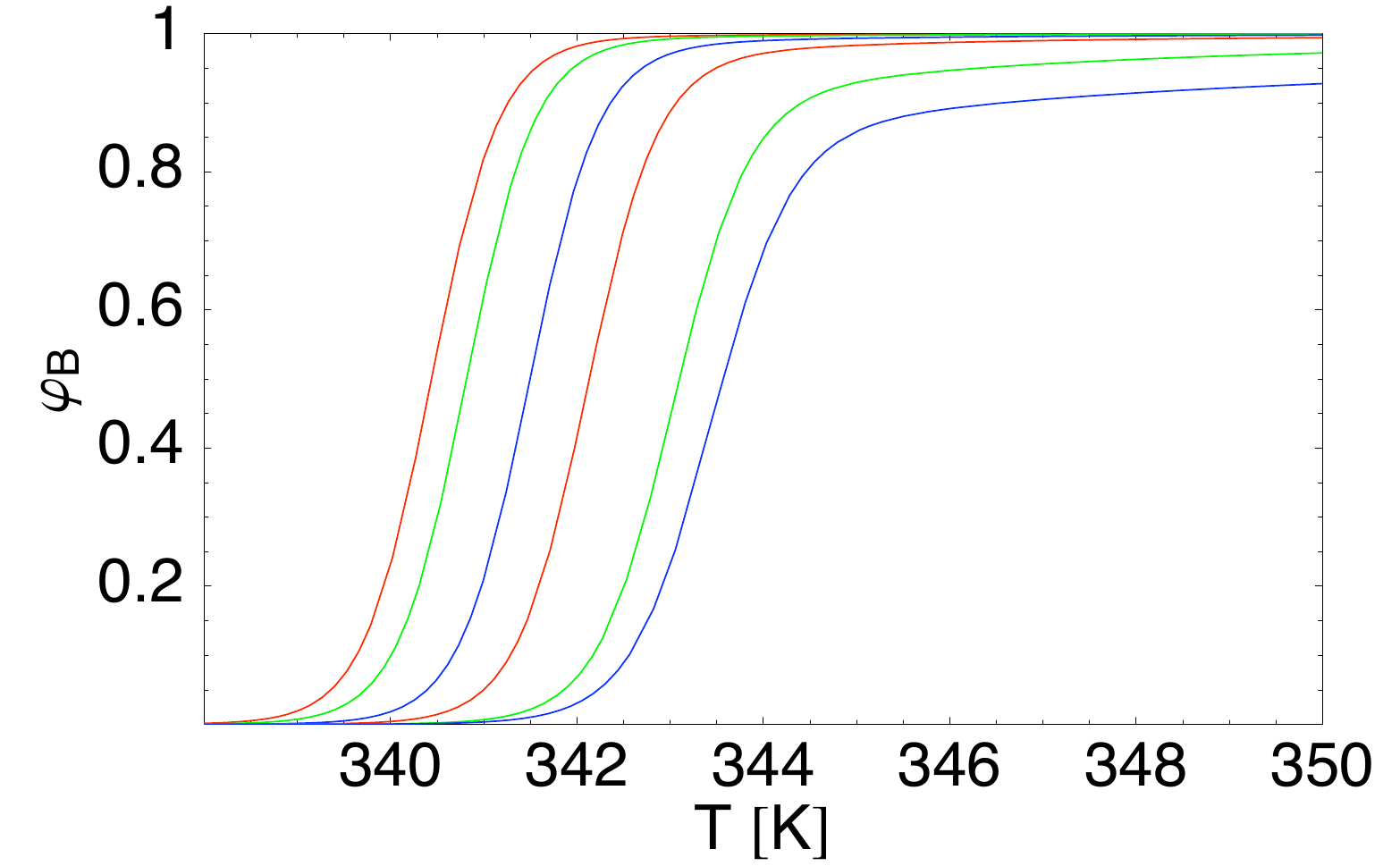}
\caption{Fraction of broken base-pairs~(\ref{phin}) vs. temperature for
$N=136$ and $f=0$ ($\mu = L = 9.87$~kJ/mol) as function of $\mu'$:
from left to right, free boundary conditions $\mu' = \mu_0$;
$\mu'/L =$ 0.86; 1.14, 1.43; 2.00; 8.56; for closed boundary
conditions the result is superimposed on the right-hand curve
($\mu'/L =+\infty$).}\label{fig4}
\end{center}
\end{figure}

\subsubsection{DNA Inserts with closed boundary conditions}

For an  A-T insert of length $N$  in more stable G-C domains a
simple starting approximation is to apply closed boundary conditions
(i.e., base-pairs $i=1$ and $N$ are considered to be held closed due
to their coupling to the adjacent G-C domains). For closed boundary
conditions,  $\tilde\mu' \rightarrow \infty$, leading to
\begin{equation} \label{RV1}
R_{V,\rm{cl}}  = \frac{\sqrt{1-\langle
c\rangle_\infty}}{\sqrt{1+\langle c\rangle_\infty}}
\end{equation}
which is non-zero for all $T>0$, implying that in this case $T^* =
0$.

Unfortunately in this case only $N-2$ base-pairs can open. A better
approach involves artificially extending the insert length from $N$
to $N+2$ and using closed boundary conditions on the extended chain.
In this case the ``fictitious" ($i=1$ and $i=N+2$) base-pairs are
held closed by the boundary conditions in order to simulate the
influence of the adjacent more stable G-C rich domains and the
remaining $N$ base-pairs can fluctuate. Since the  $i=2$ and $i=N+1$
base-pairs are adjacent to closed base-pairs their probability of
opening will be lower than that of interior ones. It is clear that
in this case melting will begin near the center of the insert.  If
$\varphi_B^{\rm cl} (N, T)$ is the fraction of open base-pairs for a
chain of length $N$ with closed boundary conditions, then simple
counting shows that the average fraction of open base-pairs in the
extended model is given by
\begin{equation} \label{phiext}
\varphi_{B}^{\rm ext} (N, T) = \frac{N+2}{N} \varphi_B^{\rm cl}
(N+2, T).
\end{equation}

A more sophisticated approach is to keep the physical insert length
of $N$ and account for the coupling to the more stable G-C rich
domains \emph{via} a mean-field type approximation by taking $ \mu_0
< \mu' < \infty$. The approaches presented above are obviously valid
only when the temperature is sufficiently far below the melting
temperature of the G-C rich domains so that the experimental UV
absorbance used to measure $\varphi_B (N, T)$ comes primarily from
the A-T inserts in the temperature range of interest.

In Fig.~\ref{fig4} we show how $\varphi_B (N, T, \tilde\mu')$ varies as a
function of $\tilde\mu'$ for $N=136$. The  melting temperature as a
function of $\tilde\mu'$ interpolates smoothly between the results
for free ($\tilde\mu' = \tilde\mu_0$) and closed boundary conditions
over a temperature range of $\sim 5$~K and the width of the
transition increases slightly with increasing $\tilde\mu'$.

\begin{figure}[ht]
\begin{center}
\includegraphics[height=4cm]{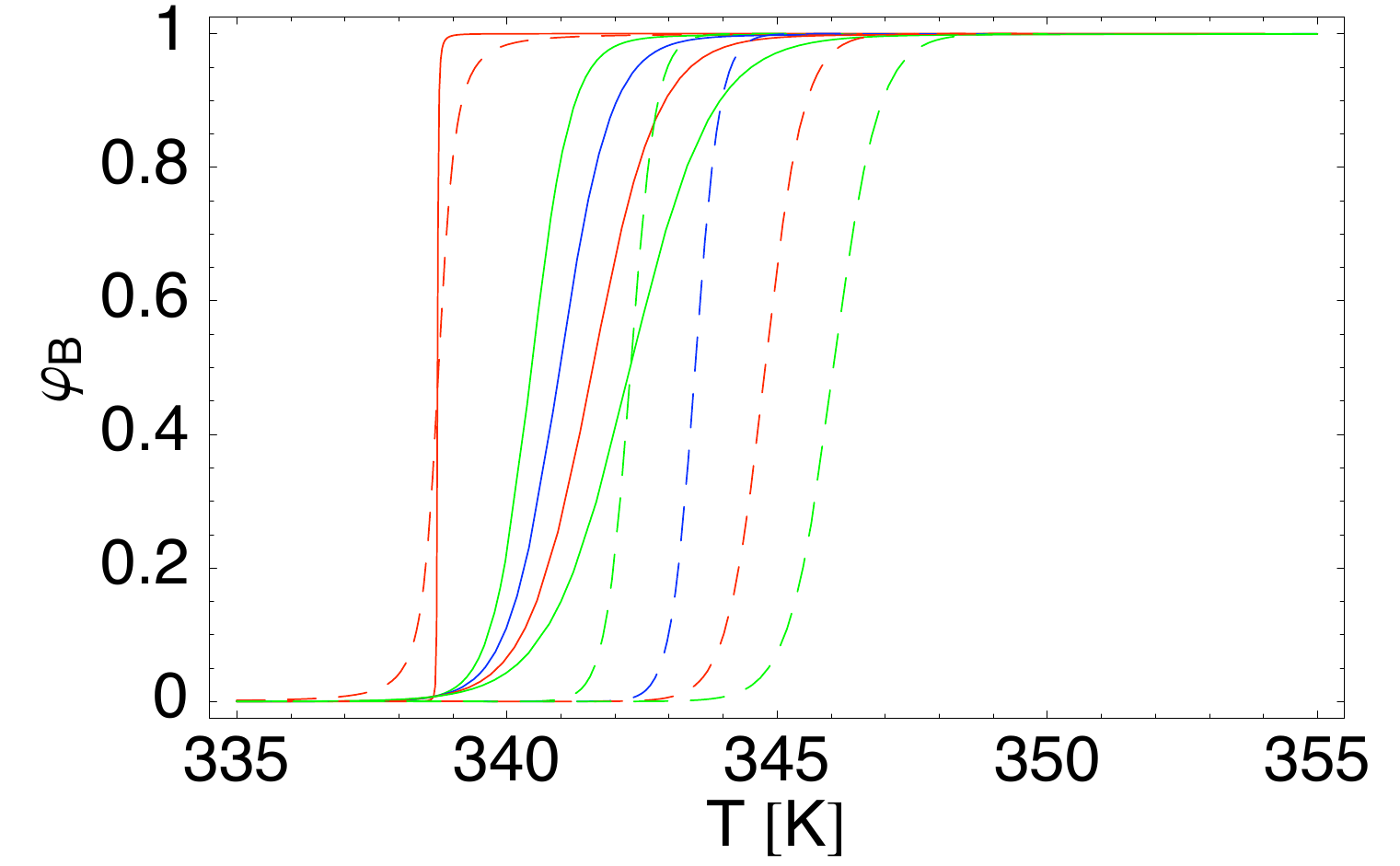}(a)
\includegraphics[height=4cm]{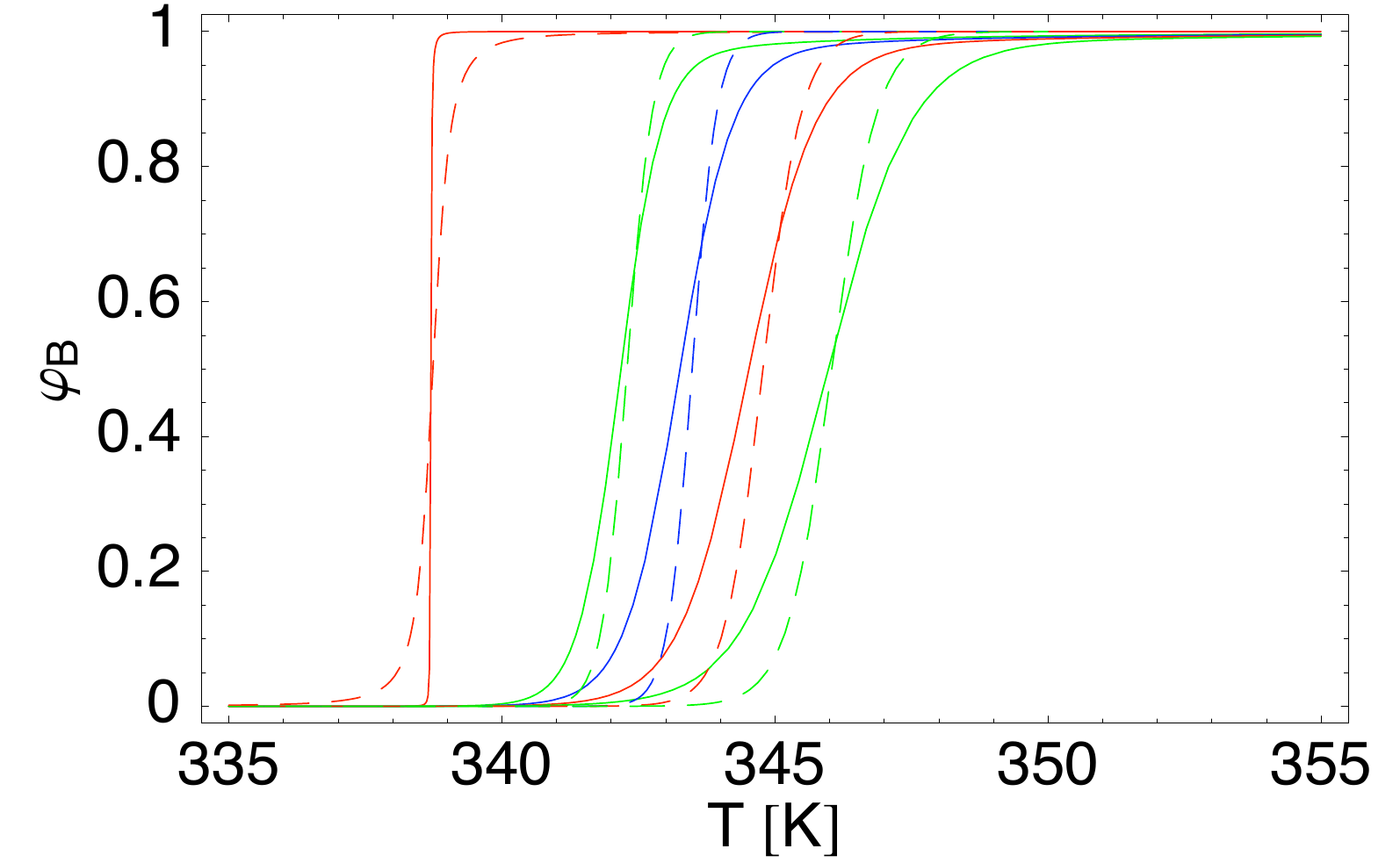}(b)
\includegraphics[height=4cm]{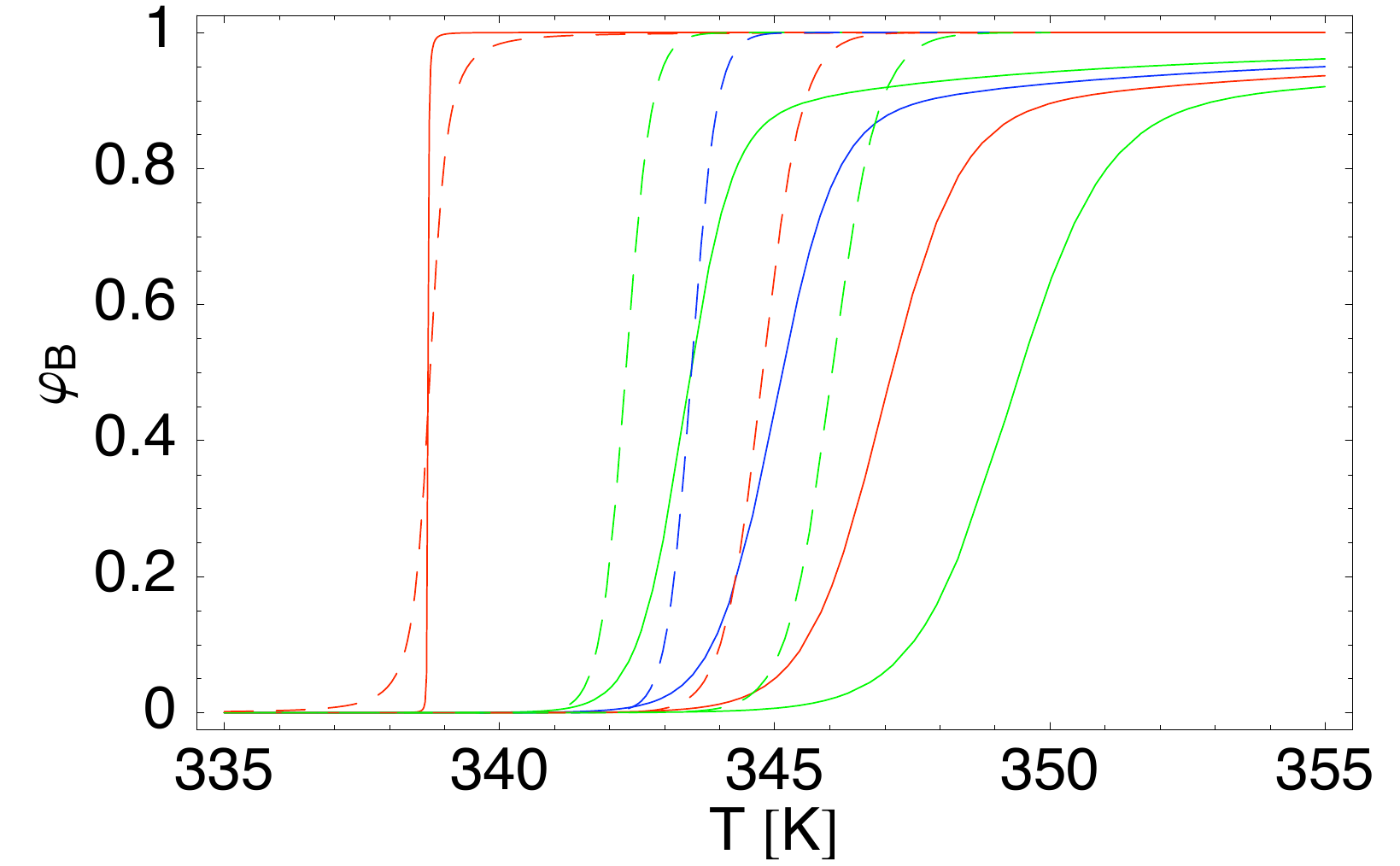}(c)
\caption{Fraction of broken base-pairs~(\ref{phin}) vs. temperature
for (from left to right) $N=30000$, 136, 105, 83, and 67: fitted experimental results
from Fig.~A.2a (dashed curves) and (a) model predictions using free
boundary conditions, (b) $\mu'$ optimized to fit $T_m(N)$, and (c)
closed boundary conditions.} \label{fig5}
\end{center}
\end{figure}

In Fig.~\ref{fig5} we compare the experimental results for A-T inserts
(Fig.~A.2a)  with the model predictions for
$\varphi_B(T,N)$  for $f=0$ and three different
model boundary conditions: (i) free boundary conditions, (ii)
optimized $\mu'$, (iii) extended model, closed boundary conditions.
The value of $L =  9.87$~kJ/mol  is held fixed to reproduce the
experimental melting temperature for $N$ = 30000 and the model
predictions for the optimized [for $T_m(N)$] $\mu'$ case are
practically insensitive to changes in $f$ and $J$. For closed
boundary conditions $\varphi_B (N)$ increases with increasing $N$ at
fixed $T$ simply because the end effects get attenuated for long
chains as illustrated in Fig.~\ref{fig5}. We  conclude
that the model in its present form can reproduce the qualitative
tendencies, but not the quantitative details, of experiments on
short A-T inserts (for such short chains including loop entropy into
the model will not lead to better fits, see below). The results
presented here do  allow us, however, to gauge the importance of
chain boundary conditions on the melting curves. One difficulty in
applying the present model arises because the simplified approach presented
here does not account for the increased probability of opening for
G-C base-pairs adjacent to the A-T inserts. The complete solution of
our model for the full heterogenous chain is in principal possible
using known numerical methods, as is discussed in the Conclusion.

The  exact result for $\varphi_B(N, T; \mu')$~(\ref{phin}) does not
reveal in a physically transparent way which states contribute the
most for a given chain length $N$ and temperature $T$ and, as
already mentioned,   includes neither the effects of loop entropy,
nor of chain sliding. In order to include such effects in a straightforward way we now study the
one-sequence approximation to the exact partition function for our
model, an approximation that should be  valid for sufficiently short chains.

\subsection{One-sequence approximation}

\subsubsection{One-sequence approximation for closed boundary conditions: DNA Inserts}

We start by examining the one-sequence approximation for homopolymer
inserts of length $N$ for closed boundary conditions without loop
entropy. The effective free energy of creating an interior
$n$-bubble with two base-pair domain walls is~\cite{JMN2},
\begin{equation} \label{Gint}
\beta\Delta G_{{\rm{int}}}^{(n)} = 4 \tilde{J}_0+2 n  \tilde{L}_0
\end{equation}
and therefore the restricted partition function, $Z_{1seq}^{\rm
cl}$, including only
 $n$-bubbles  varying in size between $n=0$ (helical insert) and $N$
 (bubble insert) is given by:
\begin{equation} \label{z1seq}
Z_{1seq}^{\rm cl} =  1 + \sum_{m = 0}^{N-1} (m+1) \exp \left[ -\beta
\Delta G_{\rm int}^{(N-m)} \right]
\end{equation}
where the first term equal to one comes from completely closed chain
state and for an $n$-bubble $m = N-n$ is the number of remaining
intact base-pairs in the insert. The factor of $(m+1) = N-n+1$ in
the sum is entropic in nature and equal to the number of ways of
placing an $n$-bubble inside an insert of length $N$.  We recall
that $L_0$ becomes negative for $T > T_m^\infty$ and therefore in
the high temperature range of interest for inserts the term
depending on $\Delta G_{{\rm{int}}}^{(n)}$  in (\ref{z1seq}) favors
large bubbles. The entropic factor, on the other hand, favors small
bubbles. The one-sequence approximation incorporates the first two
terms (of order 0 and 1) in an expansion in powers of the  loop
initiation factor,
\begin{equation} \label{lif}
\sigma_{\rm LI} = e^{-4 \tilde{J}_0},
\end{equation}
which counts the  number of bubbles~\cite{PS, PS1}.

Within the one-sequence approximation the average fraction of broken
base-pairs can be obtained from $Z_{1seq}^{\rm cl}$:
\begin{equation} \label{phid}
\varphi_{B,1seq}^{\rm cl} (N) = -\frac{1}{2N} \frac{\partial \left(
\ln Z_{1seq}^{\rm cl} \right)}{\partial L_0}
\end{equation}
The sums in~(\ref{z1seq}) can be carried out to find the following
compact expression:
\begin{equation} \label{z1ss}
Z_{1seq}^{\rm cl} =  1 + e^{-4 \tilde{J}_0}  \mathcal{C}(e^{2 \tilde{L}_0})
\end{equation}
where
\begin{equation} \label{cft}
\mathcal{C}(x) \equiv x^{-N} \left( x p'(x) + p(x) \right)
\end{equation}
with
\begin{equation} \label{pft}
p(x) \equiv  \frac{x^N-1}{x-1}
\end{equation}
By using~(\ref{z1ss}) the following expression can be obtained for
$\varphi_{B,1seq}^{\rm cl} (N)$:
\begin{equation} \label{phide}
\varphi_{B,1seq}^{\rm cl} (N) = -\frac{e^{-4 \tilde{J}_0}}{N}  \left[ \left(
\frac{\partial \mathcal{C}}{\partial x} \right) x \right]_{x=e^{
2\tilde{L}_0}}
\end{equation}
For sufficiently short chains the one-sequence approach without loop
entropy defined above will be  an accurate approximation  to the exact
result for the extended model $\varphi_{B}^{\rm ext}$ given in~(\ref{phiext}) ($N+2$ base-pairs with closed boundary conditions). When this approximation is valid,  multi-bubble states are extremely rare [the range of validity in $N$ of the one-sequence  approximation depends on the value of  $J_0$ via $\sigma_{LI}$~(\ref{lif})].

Although  it is difficult to incorporate  bubble loop entropy into
our model in a general way because of mathematical complications
arising from the ``long-range" nature of the loop entropy factor, it
is easy to do so within the one-sequence approximation. Including
the loop entropy lowers the probability of $n$-bubble opening. We
adopt a common simplified form for the loop entropy factor associated with
$n$ broken base pairs~\cite{gotoh,fixfr, wartell},
\begin{equation} \label{fle}
g_{\rm LE}(n) = (n_0 + 2 + 2 n)^{-k}
\end{equation}
that depends on the bubble loop length, $\ell_{\rm B} = 2 + 2 n$, and
is parametrized by a constant $n_0$ and an exponent $k$. The loop
entropy exponent $k$ is thought to be in the range $3/2 \leq k \leq
2.1$, depending on the extent to which chain self-avoidance is taken
into account~\cite{peliti}. The term $n_0$ accounts for the enhanced
difficulty of forming small closed bubbles arising from DNA chain
stiffness. Including the loop entropy leads to a modified
one-sequence partition function, given by
\begin{equation} \label{z1sle}
Z_{1seq}^{\rm cl, LE} =  1 + \sum_{m = 0}^{N-1} (m+1) g_{\rm LE}(N-m) \exp \left[ -\beta \Delta G_{\rm int}^{(N-m)} \right].
\end{equation}
The introduction of loop entropy ($k>0$) in $Z_{1seq}^{\rm cl, LE}$ can
have an exaggerated  effect on the calculated melting curves if the
loop initiation factor, $\sigma_{\rm LI}$ (\ref{lif}), is not
readjusted at the same time. If we define $D = (n_0+2)/2$ and use
$\tilde{J}_0 \rightarrow  \tilde{J}_0+(k/4)\ln (2D)$ in (\ref{z1sle}) then
$Z_{1seq}^{\rm cl, LE}$ can be rewritten as
\begin{equation} \label{z1slem}
Z_{1seq}^{\rm cl, LE} =  1 + \sum_{m = 0}^{N-1} (m+1) [1 +
(N-m)/D]^{-k} \exp \left[ -\beta \Delta G_{\rm int}^{(N-m)} \right].
\end{equation}
with $G_{\rm int}$ still given by (\ref{Gint}) (in the fitting of
experimental data, the value of $D$ has been taken to be as large as
96~\cite{blake} and even 450~\cite{gotoh}). The above readjustment
of $J_0$ means that only long $n$-bubbles ($n =  N-m > D$) ``feel" the
effect of loop entropy (the suppression of short bubble formation
due to increased chain stiffness being incorporated directly into
the  readjusted $J_0$). We will compare the predictions of the
one-sequence approximation with ($k, D>0$) and without ($k=0$) loop
entropy using (\ref{z1slem}). Although the sums in~(\ref{z1slem})
apparently cannot be carried out analytically, once they are
performed numerically, the analog of~(\ref{phid}) can be used to
obtain $\varphi_{B,1seq}^{\rm cl,LE}$.

\begin{figure}[ht]
\begin{center}
\includegraphics[height=6cm]{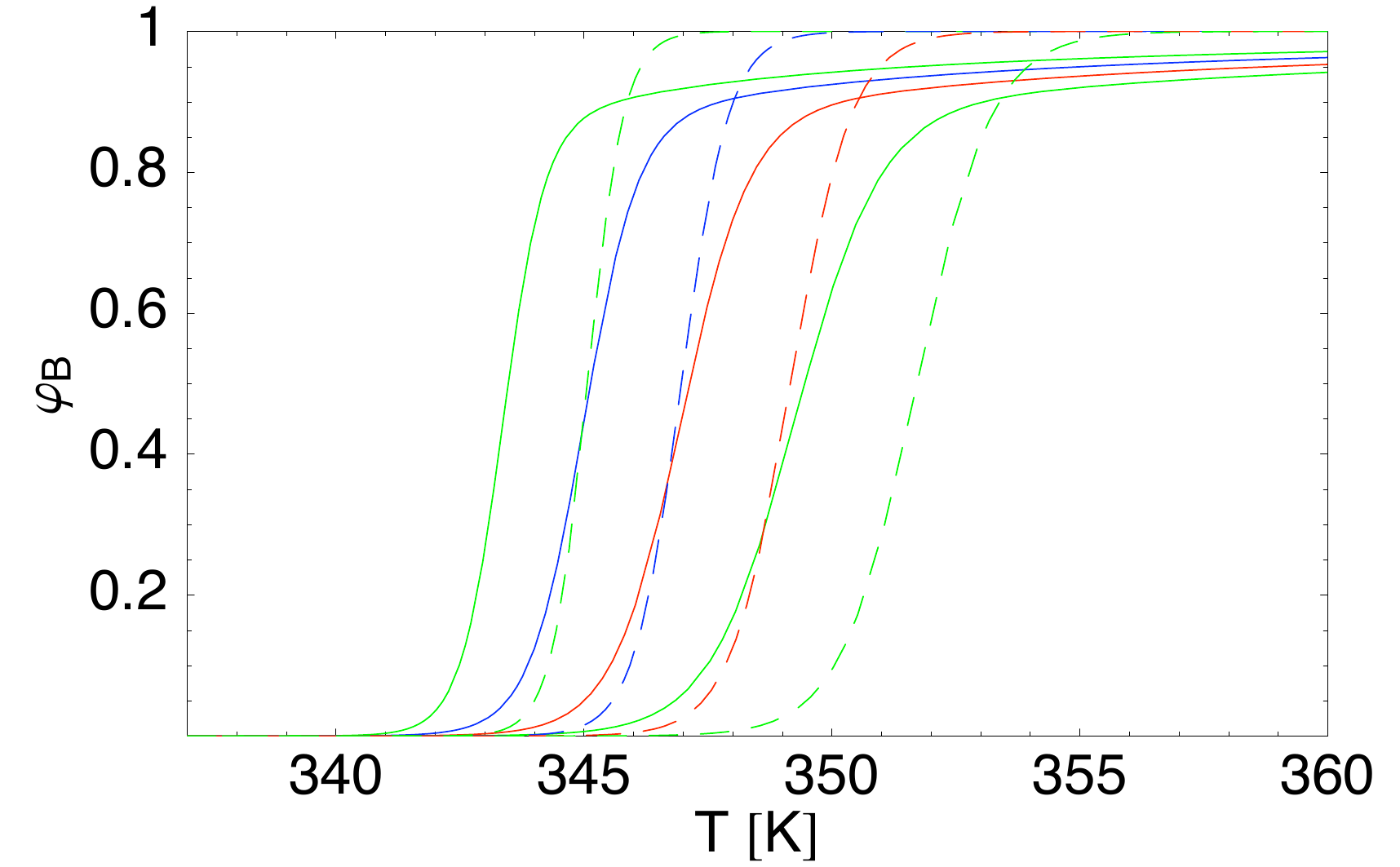}
\caption{Comparison of the two-state approximation (dashed curves)
without loop entropy~(\ref{phi2stwle}) with the full
result~(\ref{phiext}) (solid curves) for closed boundary conditions;
from left to right $N=136$, 105, 83, and 67 (same parameters and
colors as Figure~\ref{fig3}).} \label{fig6}
\end{center}
\end{figure}

If in evaluating the one-sequence partition function, $Z_{1seq}^{\rm
cl,LE}$, we retain only the completely closed ($m=N$) and completely open
($m=0$) states, we obtain the \textit{two-state approximation}:
\begin{equation} \label{phi2st}
\varphi_{B,2st}^{\rm cl, LE} = \frac{ 1}{1 +  \left\{ (1+N/D)^{-k}
\exp \left[ -\beta \Delta G_{\rm int}^{(N)} \right]\right\}^{-1}}
\end{equation}
A more general $s$-state approximation can be defined by including
the $m = 0,\ldots,s-2$ terms in the sum  (\ref{z1sle}). Without loop
entropy ($k=0$)~(\ref{phi2st}) simplifies to
\begin{equation} \label{phi2stwle}
\varphi_{B,2st}^{\rm cl}=\frac12\left\{1-\tanh\left[-\beta\Delta
G_{\rm int}^{(N)}/2 \right]\right\}
\end{equation}

In Figure~\ref{fig6}, the  2-state approximation without loop entropy
is compared to the exact result for the extended case~(\ref{phiext}). We observe a
cross-over temperature (at which the 2-state approximation begins to
overestimate $\varphi_{B}^{\rm cl}$) roughly given by the
temperature at which $\Delta G_{{\rm{int}}}^{(N)}$ goes from
positive to negative (signaling a vanishing ``nucleation barrier"
for the completely open insert). Contrary to previous
claims~\cite{blake}, in the present case the two-state approximation
overestimates $T_m(N)$ by more than 2~K and underestimates the
transition width.

The form (\ref{z1slem}) suggests defining an effective total
$n$-bubble free energy
\begin{equation} \label{freenb}
\beta \Delta F_{\rm int}^{(n)}  =  \beta \Delta G_{\rm
int}^{(n)}-\ln (N-n+1 )  + k \ln ( 1 + n/D)
\end{equation}
that accounts for the intrinsic free energy of bubble formation
(first term), as well as positional (second term) and loop entropy
(third term). $\beta \Delta G_{\rm
int}^{(n)}$ decreases with increasing $n$ for $T>T_m^{\infty}$ ($L_0<0$) and increases for $T<T_m^{\infty}$ ($L_0>0$). The positional and loop entropy contributions increase the effective free
energy cost of bubble creation as the bubble size $n$ increases.

\begin{figure}[ht]
\begin{center}
\includegraphics[height=4cm]{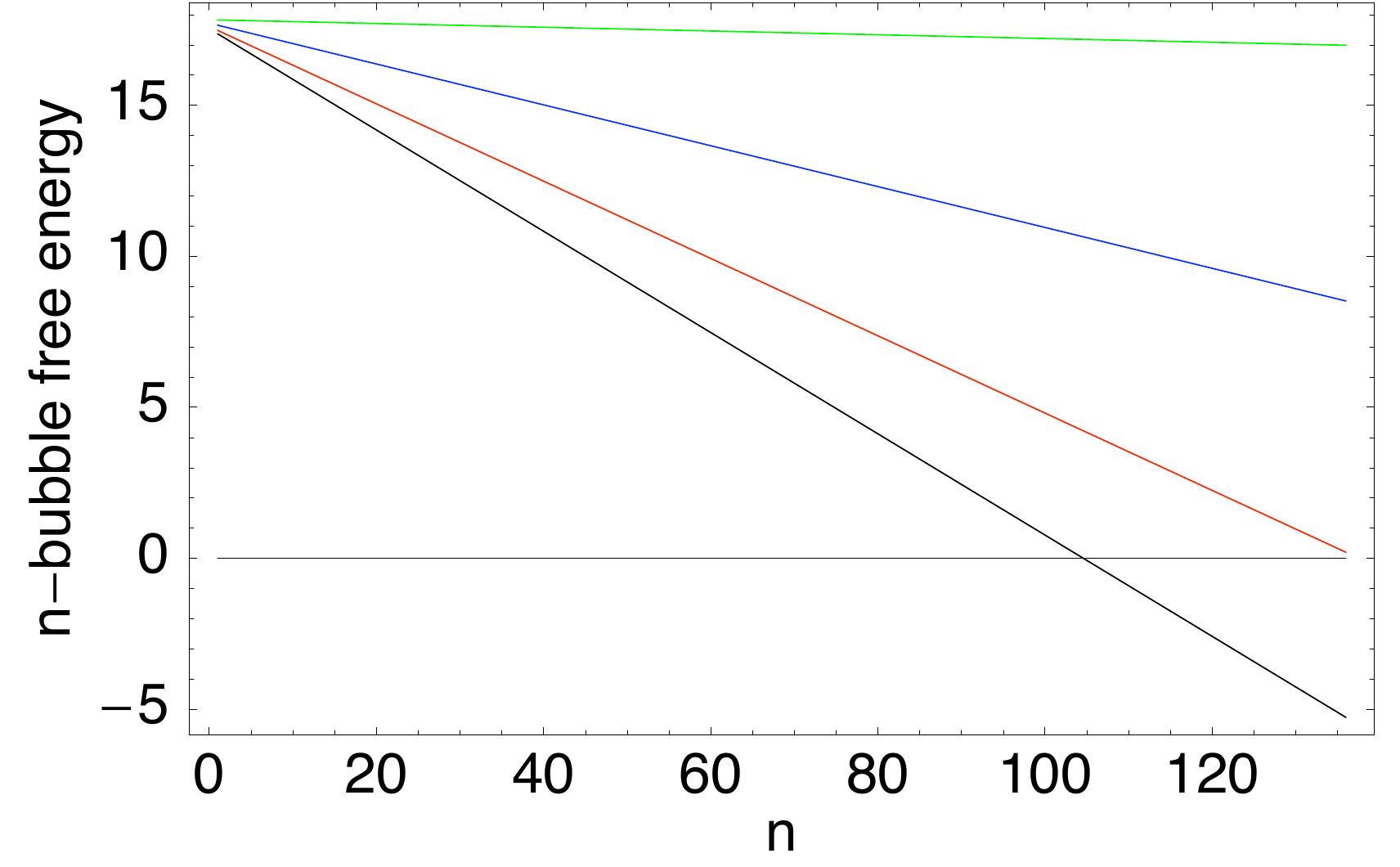}(a)
\includegraphics[height=4cm]{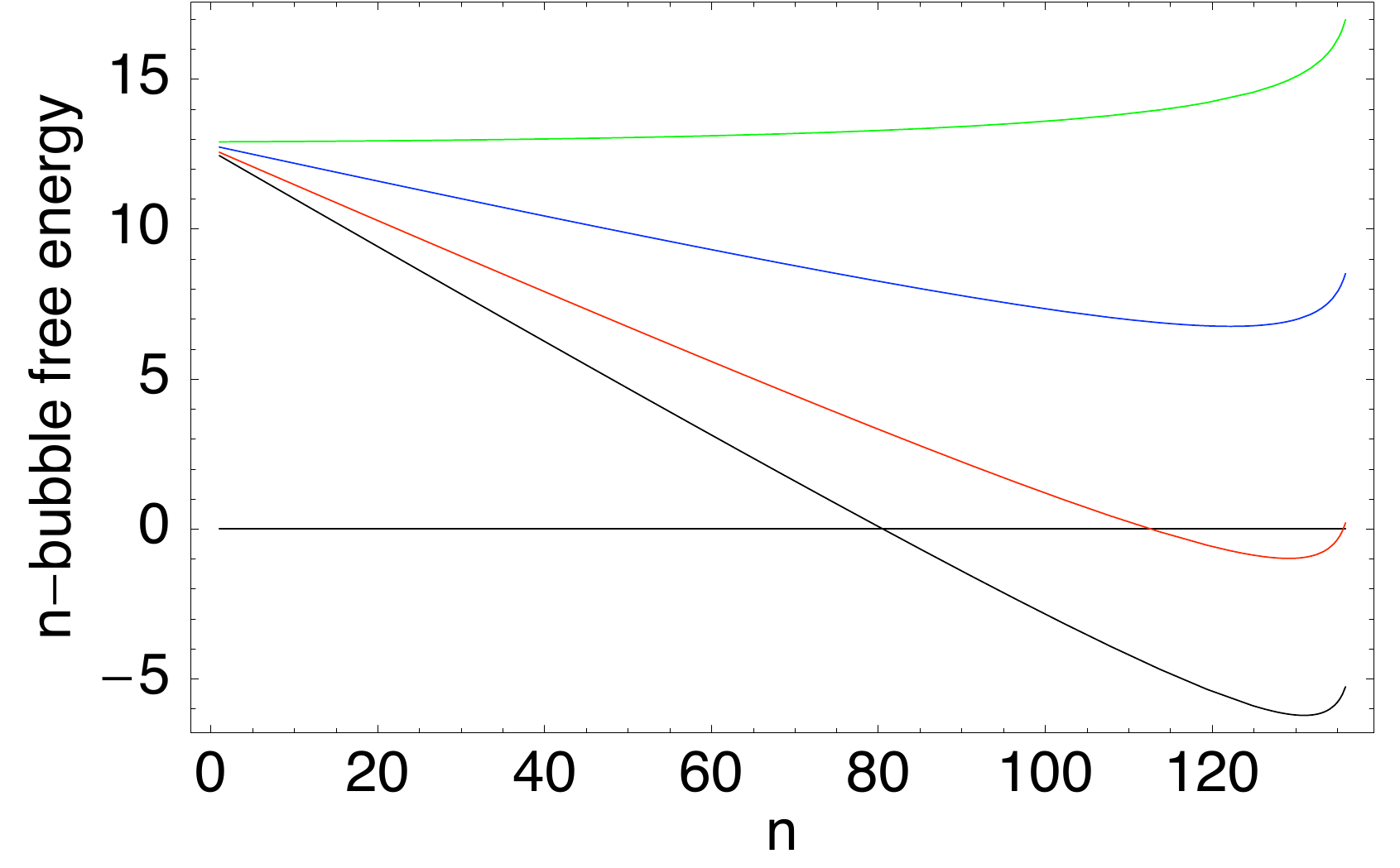}(b)
\includegraphics[height=4cm]{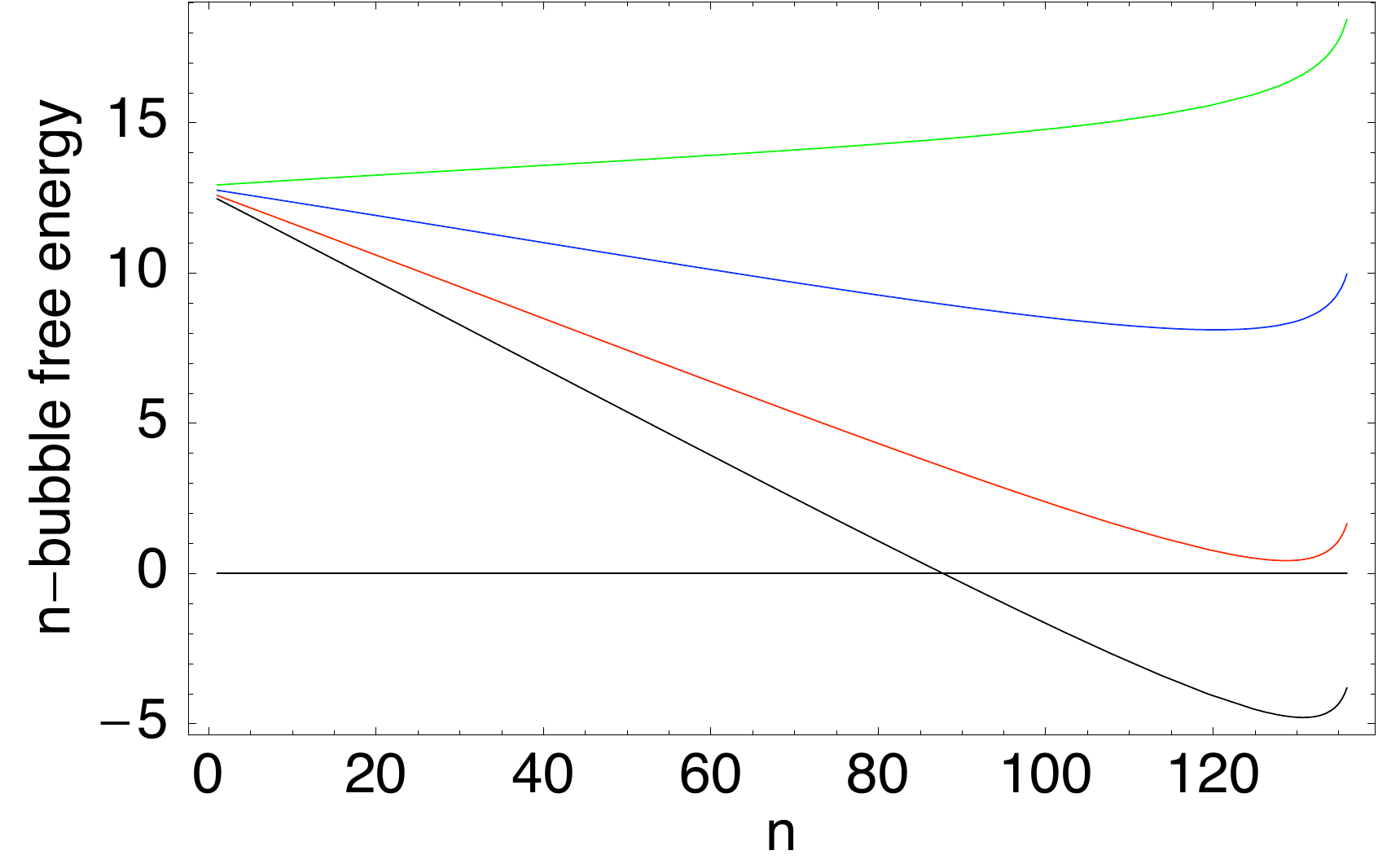}(c)
\includegraphics[height=4cm]{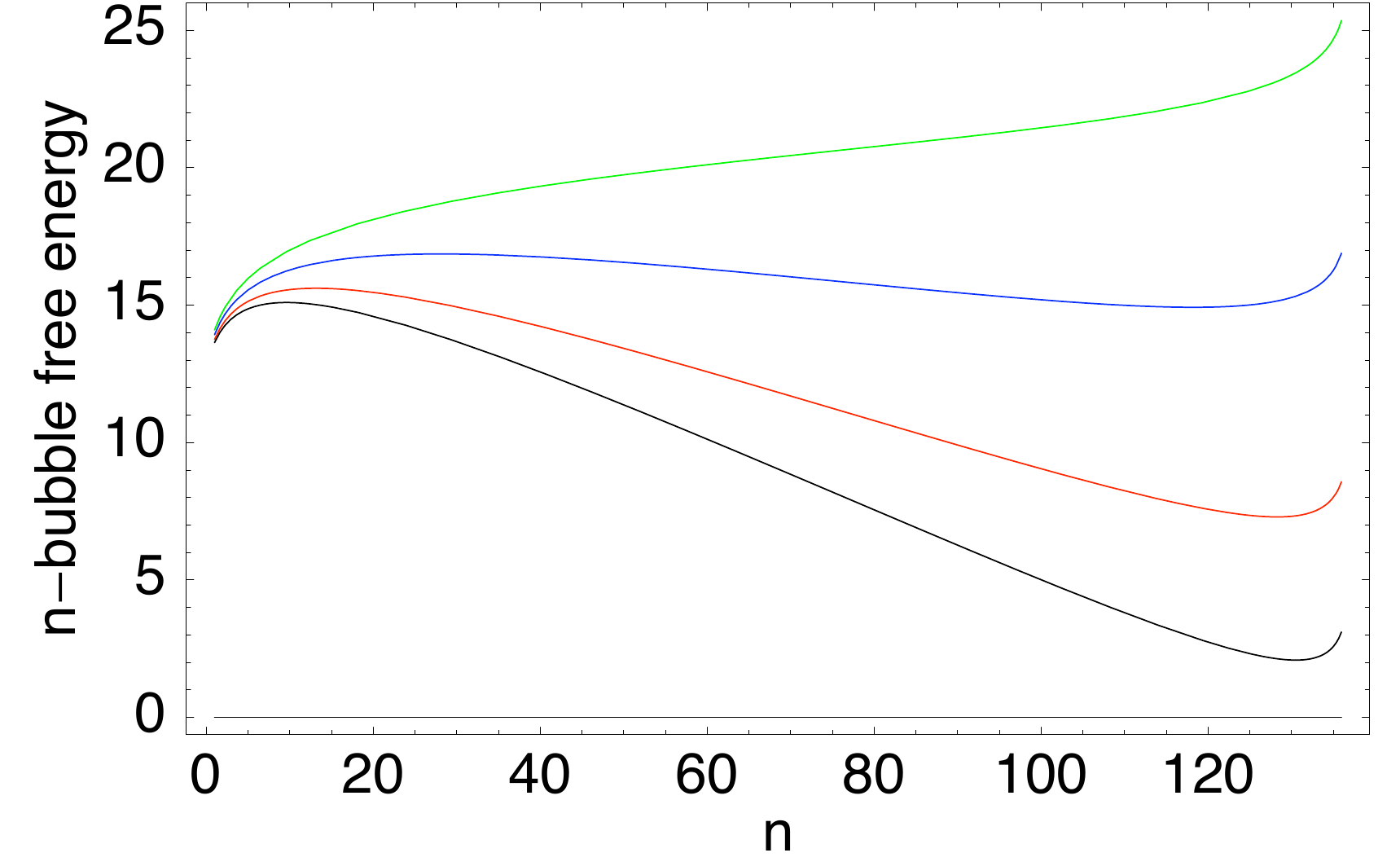}(d)
\caption{Free energy of bubble formation for $N=136$ and $T=339, 342, 345, 347$~K (from top to bottom): (a) intrinsic free energy, $\beta \Delta G_{\rm int}^{(n)}$; total free energy, $\beta\Delta F_{\rm int}^{(n)}$: (b) $k=0$ (without loop entropy); (c)
$k=1.7$, $D=100$; (d) , $k=1.7$, $D=1$ ($f=0$, other model
parameters as in Fig.~\ref{fig3}).}\label{fig7}
\end{center}
\end{figure}

In Fig.~\ref{fig7} we plot bubble free energies  for $N=136$ and increasingly
important loop entropy effects. The intrinsic part, $\Delta G_{\rm
int}^{(n)}$ is a linearly decreasing function of $n$ and vanishes at
$T=345$~K, close to  the temperature at which the 2-state approximation becomes an
overestimation (see Fig.~\ref{fig7}). We observe that (i) inclusion of the
positional entropy alone (Fig.~\ref{fig7}b) leads to a minimum in~(\ref{freenb})  near $n = N$ for sufficiently high temperatures
 and (ii) the loop entropy rigidity parameter  $D$ plays a
minor role when it is close to 100 (Fig.~\ref{fig7}c) and an important one 
when it is close to 1, the value commonly used in the modeling of
infinite chains (Fig.~\ref{fig7}d). In the latter case (\ref{freenb}) remains positive over the whole temperature range studied  and has a maximum for small $n$ and a minimum near $n = N$ for sufficiently high temperatures.

\begin{figure}[t]
\begin{center}
\includegraphics[height=6cm]{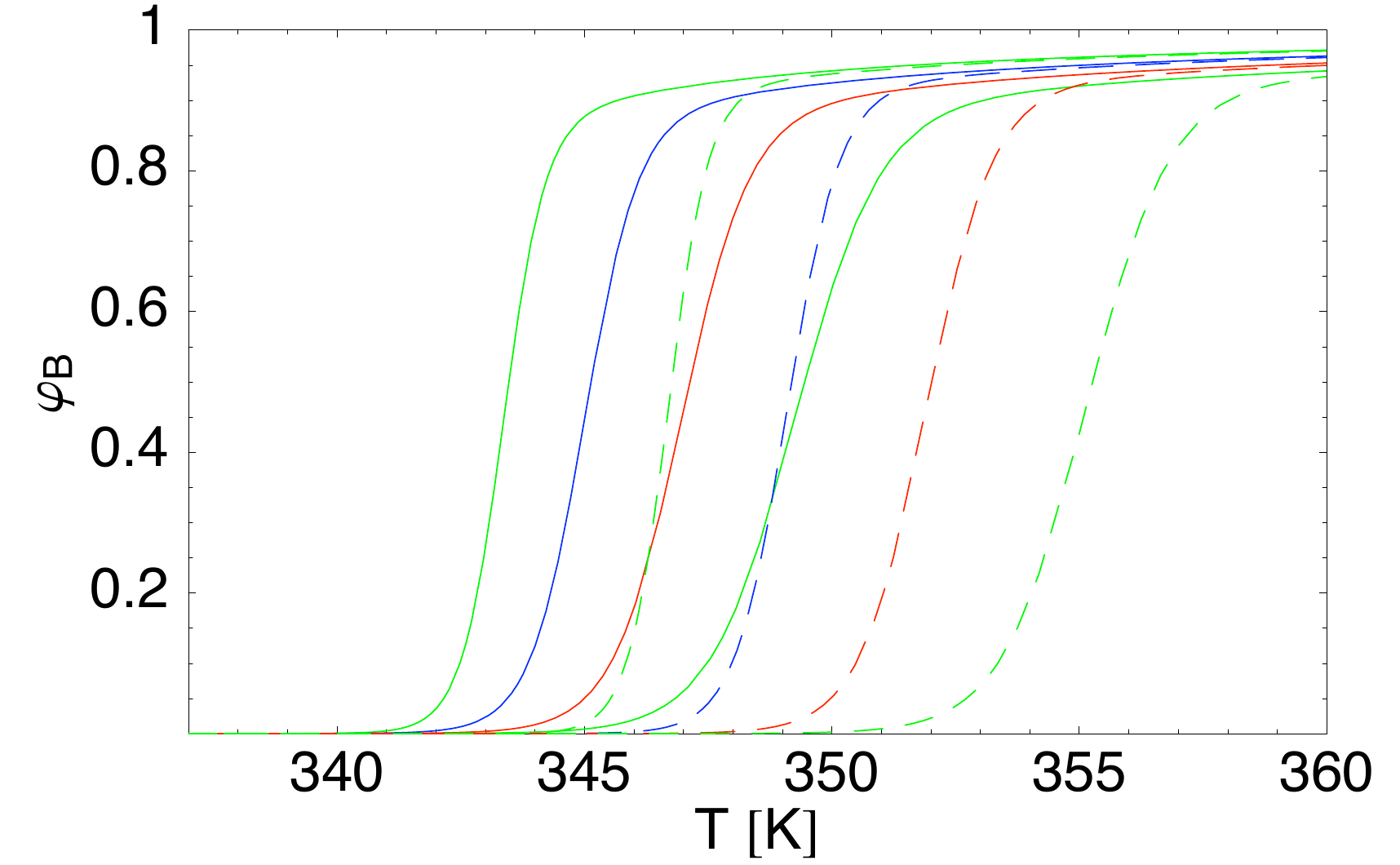}(a)\\
\includegraphics[height=6cm]{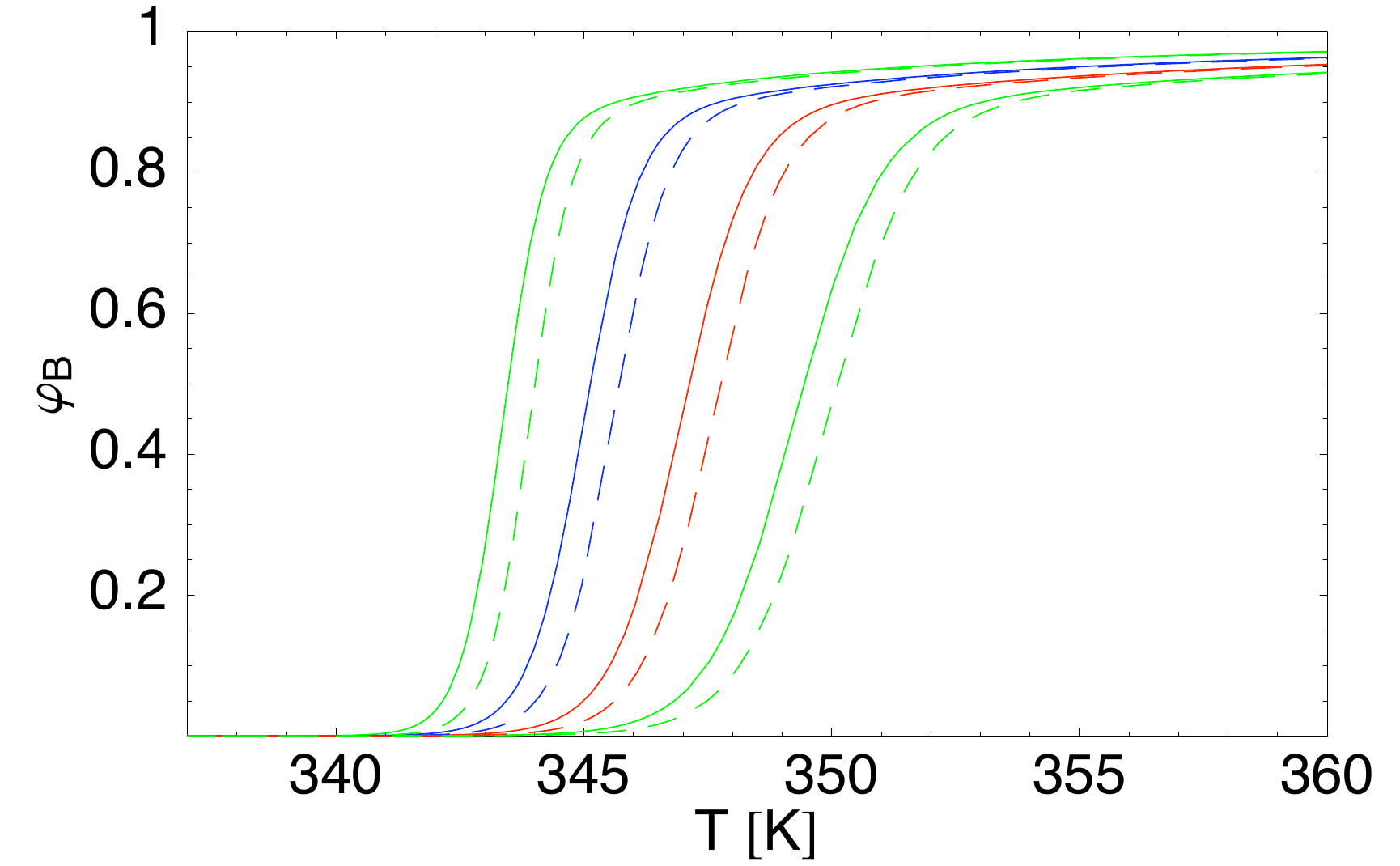}(b)
\caption{Comparison of the one-sequence approximations with loop
entropy (dashed curves) and without (solid curves)
~(\ref{z1slem}), from left to right $N=136$, 105, 83, and 67 for
closed boundary conditions (same parameters as in Figure~\ref{fig3})
and $k = 1.7$ (a)  $D = 1$ (b) $D = 100$.} \label{fig8}
\end{center}
\end{figure}

In Fig.~\ref{fig8}, we compare the one-sequence approximations
with and without loop entropy~(\ref{phi2stwle}) for short inserts
obeying closed boundary conditions. For $J=9.13$~kJ/mol we find that for inserts without loop entropy the one-sequence approximation is practically
indistinguishable from the exact result~(\ref{phiext}) for $N<10000$.
Because loop entropy further reduces the probability of bubbles, we therefore
believe that the one-sequence approximation with loop entropy should
be an excellent approximation in most cases of practical interest
(i.e.,  inserts with lengths less than a few thousand
base-pairs). We observe in Fig.~\ref{fig8} that for such inserts and fixed $L$ the net
result of including the loop entropy is to shift the melting curves
to the right by about 10~K for $D = 1$ and about 2~K for $D = 100$
without much change in the transition width. It therefore seems as
if  the addition loop entropy will not enable us to improve the fits
to experiment shown in Fig.~\ref{fig5}.

Although it is possible to work out the details of the one-sequence
approximation when the end base-pairs in an insert of length
$N$ experience a chemical potential $\mu'<\infty$, we will not
present these results here.

\subsubsection{One-sequence approximation for free boundary conditions}

\begin{figure}[t]
\begin{center}
\includegraphics[height=2.5cm]{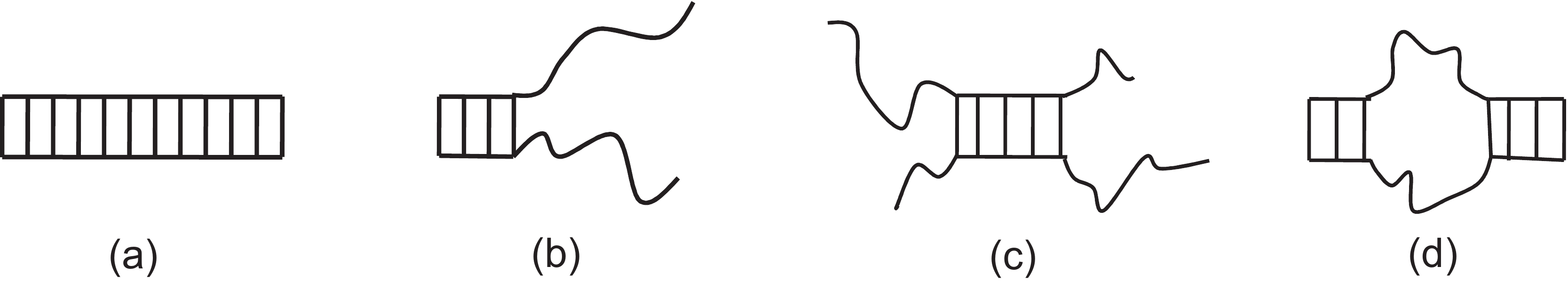}
\caption{The four DNA states accounted for in the one-sequence approximation for free polymers (aside from the dissociated chains): (a) closed chain, (b) end unwinding (c) internal helix (d) internal bubble, corresponding, respectively to the first four terms in \ref{z1sop}.}\label{fig9}
\end{center}
\end{figure}

We now examine the one-sequence approximation with and without loop
entropy for DNA homopolymers
of length $N$ with free boundary conditions. Because most synthetic DNA homopolymers are less than a few
thousand base-pairs long~\cite{PS, PS1, blake, montroll},
the one-sequence approximation may  be a useful and accurate simplified
approach in such cases. For free boundary conditions, besides single interior
bubbles, we must include the possibility of single helical
sequences. The effective free energy of creating an interior
$n$-bubble with two base-pair domain walls is given in~(\ref{Gint});
the effective free energy of creating a single unzipped sequence of
length $n$ starting at $i=1$ or $i=N$ (with only one base-pair
domain wall) is~\cite{JMN2}:
\begin{equation}
\beta \Delta G_{{\rm{end}}}^{(n)} = 2\tilde{J}_0  -  \tilde{K}_0  + 2 n \tilde{L}_0
\label{gend}
\end{equation}
The effective free energy for creating a single interior helical
sequence  of length $m = N-m$ (including neither the $i=1$ or $i=N$
base-pair) with two domain walls is~\cite{JMN2}:
\begin{equation}
\beta \Delta G_{{\rm{helix}}}^{(m)} = 4\tilde{J}_0  -  2\tilde{K}_0  + 2 (N-m) \tilde{L}_0
\label{ghelix}
\end{equation}
The effective free energy needed to completely denature the DNA
chain of length $N$ is $\beta \Delta G_{\rm open}^{(N)} = 2 \tilde{L}_0 N -
2 \tilde{K}_0$. The restricted one-sequence partition function for free
boundary conditions, $Z_{1seq}^{\rm free}$, includes contributions
from (i) the completely closed state (dsDNA), normalized to a weight
of one, (ii) interior $n$-bubbles inserted in a domain of length
$N-2$ varying in size between $n=1$ and $N-2$,
\begin{equation} \label{z1sint}
Z_{1seq}^{\rm B_{int}} = \sum_{m = 0}^{N-3} (m+1) \exp \left[ -\beta
\Delta G_{\rm int}^{(N-2-m)} \right]
\end{equation}
(iii) one unzipped end sequence of length $n$, $Z_{1seq}^{\rm end}$
with two-fold degeneracy
\begin{equation} \label{z1send}
Z_{1seq}^{\rm end}=2\sum_{n = 1}^{N-1} \exp\left[-\beta \Delta
G_{\rm end}^{(n)} \right]
\end{equation}
(iv) a single interior Helical sequence
\begin{equation} \label{z1sHint}
Z_{1seq}^{\rm H_{int}} = \sum_{m = 1}^{N-2} (N-1-m)^\varepsilon \exp
\left[ -\beta \Delta G_{\rm helix}^{(N-2-m)} \right],
\end{equation}
where $\varepsilon = 1$  without chain sliding (for heteropolymers using
average parameter values) and 2 with (for homopolymers like
polydA-polydT) \cite{PS, PS1, montroll}, (iv) the completely open
state (op),
\begin{equation} \label{z1scop}
Z_{1seq}^{\rm op} =  \exp \left[ -\beta \Delta G_{\rm
open}^{(N)}\right].
\end{equation}
$Z_{1seq}^{\rm free}$ can therefore be written as
\begin{equation} \label{z1sop}
Z_{1seq}^{\rm free}= 1 + Z_{1seq}^{\rm
end}+Z_{1seq}^{\rm H_{int}} +Z_{1seq}^{\rm B_{int}}+Z_{1seq}^{\rm op}
\end{equation}
The four DNA states accounted for in the one-sequence approximation (aside from the dissociated chains) are shown in Fig.~\ref{fig9}.

It is now easy to include loop entropy by inserting the loop entropy
factor $g_{\rm LE}$ into the second term of~(\ref{z1sop}):
\begin{equation} \label{z1sintle}
Z_{1seq}^{\rm B_{int}, LE} = \sum_{m = 0}^{N-3} (m+1)[n_0 +2 + 2
(N-m)]^{-k} \exp \left[ -\beta \Delta G_{\rm int}^{(N-2-m)} \right]
\end{equation}

It is not possible now to simply readjust $J_0$ as was done
for inserts, because unzipped end sequences ``see" the un-readjusted
$J_0$. Unzipped end sequences are composed of two unbound chains
joined at one end and therefore there is  no loop entropy factor in
$Z_{1seq}^{\rm end}$ or $Z_{1seq}^{\rm H_{int}}$ (a small correction
term for two such self-avoiding chains, however, has been neglected,
see~\cite{garorl}). We can, however, rewrite (\ref{z1sintle}) as
\begin{equation} \label{z1sintlem}
Z_{1seq}^{\rm B_{int}, LE} = \sum_{m = 0}^{N-3} (m+1)[1+
(N-m)/D]^{-k} \exp \left[ -\beta \Delta \hat{G}_{\rm int}^{(N-2-m)} \right],
\end{equation}
$\beta\Delta \hat{G}_{\rm int}^{(n)}$ is the same as $\beta\Delta G_{\rm int}^{(n)}$ with $\tilde{J}$ replaced by
\begin{equation} \label{jp}
 \hat{J} \equiv   \tilde{J}+(k/4)\ln (2D) > \tilde{J}.
\end{equation}
It is then  possible to define an effective loop initiation factor,
$\hat{\sigma}_{\rm LI} \equiv e^{-4 \hat{J}_0} <\sigma_{\rm LI}$, that controls the probability of bubble formation in the presence of loop entropy and depends on the readjusted value  $\hat{J}$ (although it is still $\sigma_{\rm LI}$ that controls the probability of end unwinding and one internal helical section).

Within the free boundary condition one-sequence approximation the
average fraction of broken base-pairs can be obtained from
$Z_{1seq}^{\rm op}$ \textit{via}
\begin{equation} \label{phiopd}
\varphi_{B,1seq}^{\rm free}(N)=-\frac{1}{2N} \frac{\partial \left(
\ln Z_{1seq}^{\rm free} \right)}{\partial L_0}
\end{equation}

When chain dissociation is taken into account the contribution from
the completely open chain, $Z_{1seq}^{\rm op}$, is dropped from
$Z_{1seq}^{\rm free}$, which then becomes the internal partition function
for associated chains:
\begin{equation} \label{z1sopass}
Z_{1seq}^{\rm free}= 1 + Z_{1seq}^{\rm end}+Z_{1seq}^{\rm
H_{int}}+Z_{1seq}^{\rm B_{int}} \quad \quad \mbox{\rm (associated
chains)}.
\end{equation}
The corresponding $\varphi_{B,1seq}^{\rm free}$ is the fraction of
broken base-pairs in associated chains (clearly a lower bound for
the experimentally measured total fraction of broken base-pairs,
because the contribution of dissociated chains is neglected). In
this case the one-sequence approximation~(\ref{z1sopass}) incorporates the first
four terms (of order 0, 1/2 and 1 for the last two terms) in an expansion in powers of
the loop initiation factor, $\sigma_{\rm LI}$ [the so-called zipper
model neglects the last (bubble) contribution]~\cite{PS, PS1}. The
next higher term, neglected in (\ref{z1sop}) and of order 3/2, accounts
for one internal bubble with chain sliding. In most cases of
practical interest there is little difference between using
(\ref{z1sop}) and (\ref{z1sopass}).

\begin{figure}[ht]
\begin{center}
\includegraphics[height=4cm]{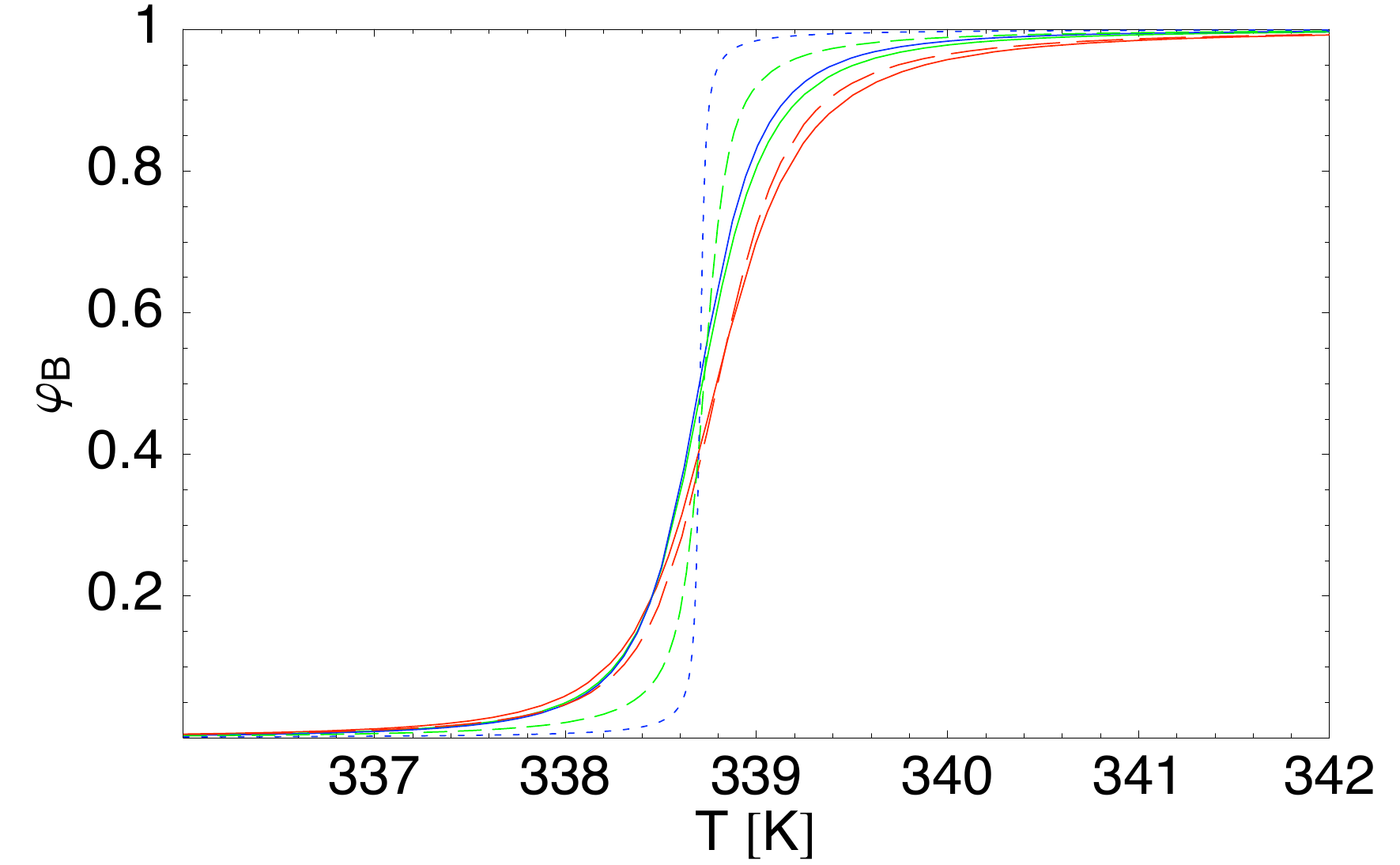}(a)
\includegraphics[height=4cm]{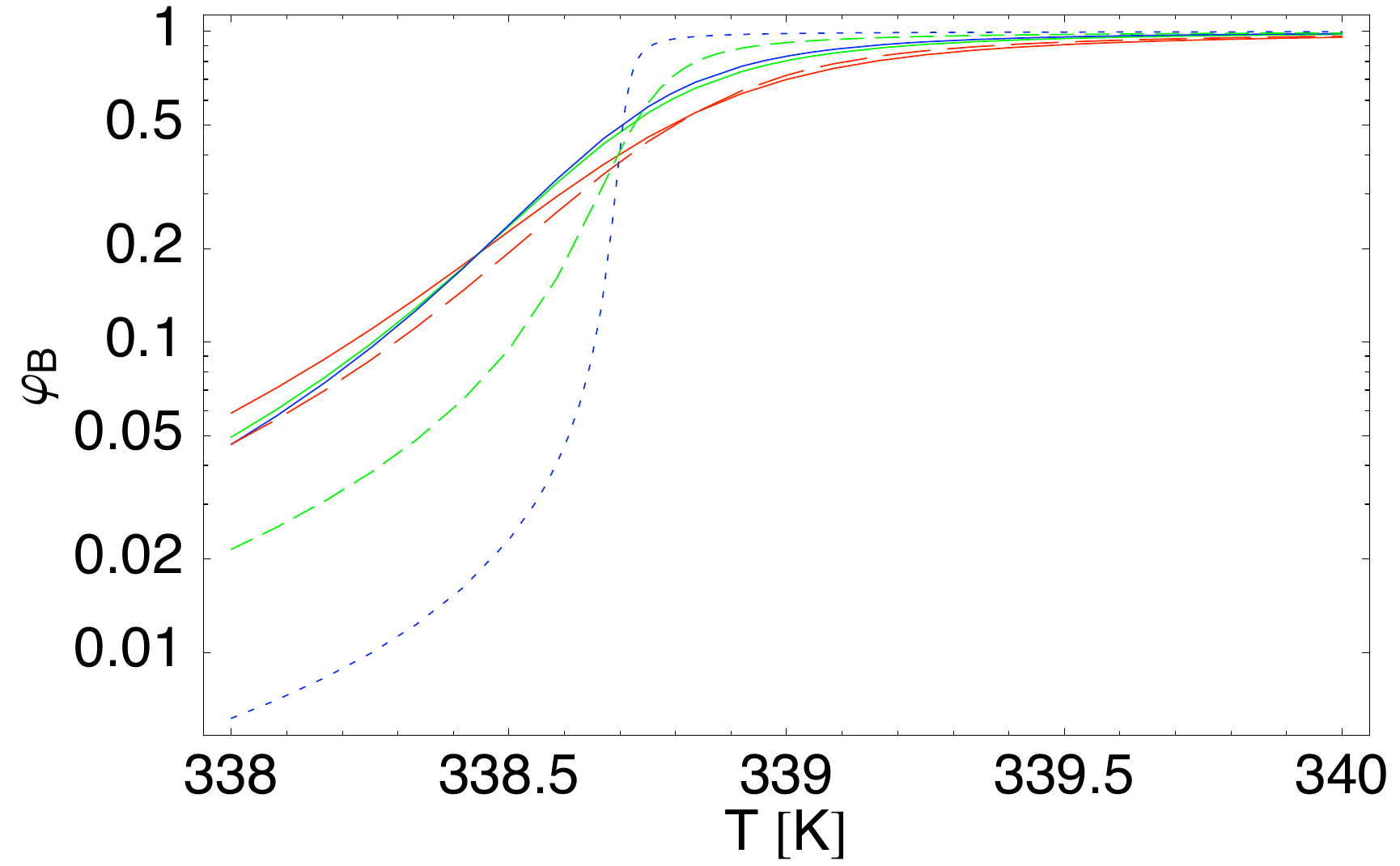}(b)
\includegraphics[height=4cm]{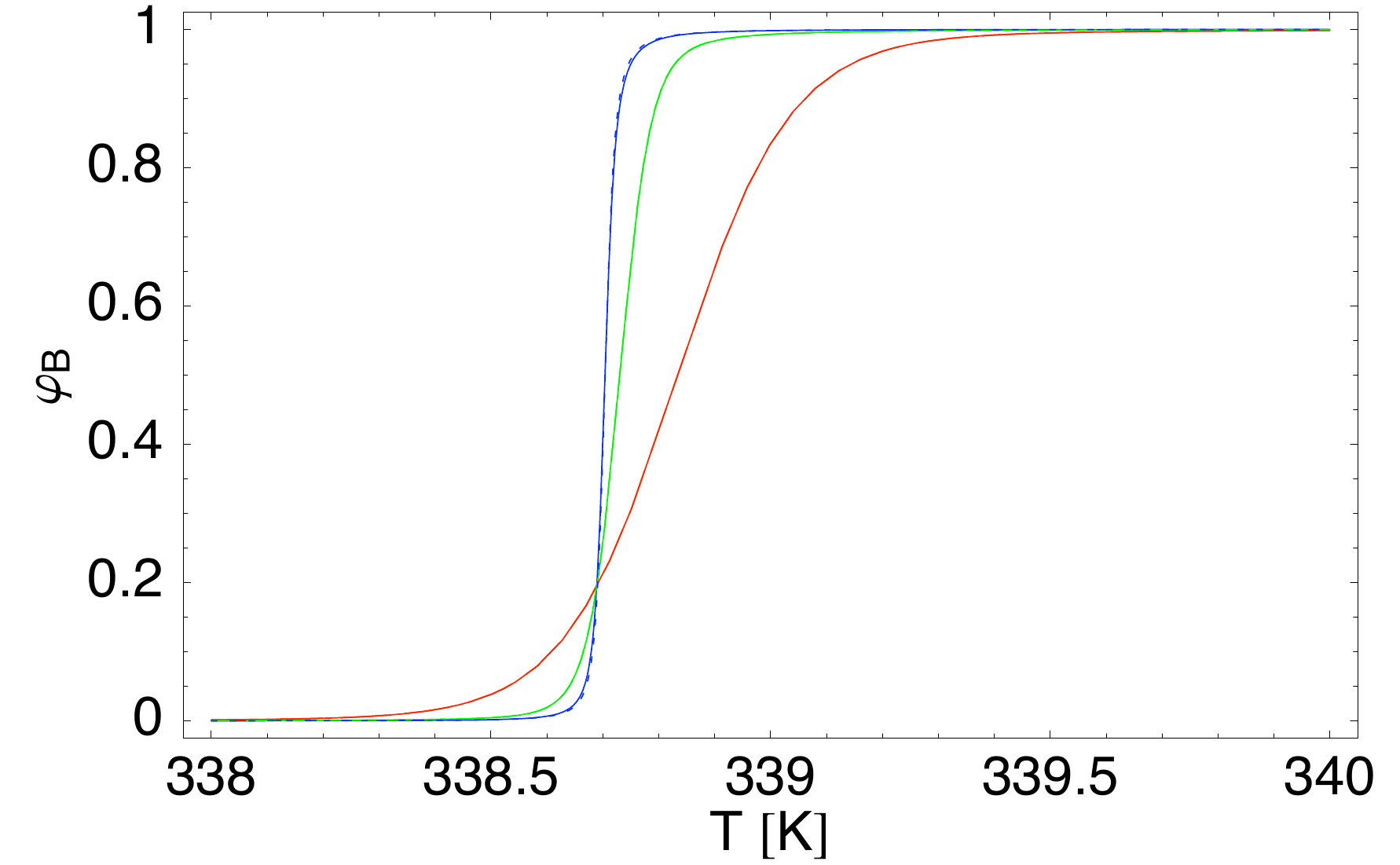}(c)
\includegraphics[height=4cm]{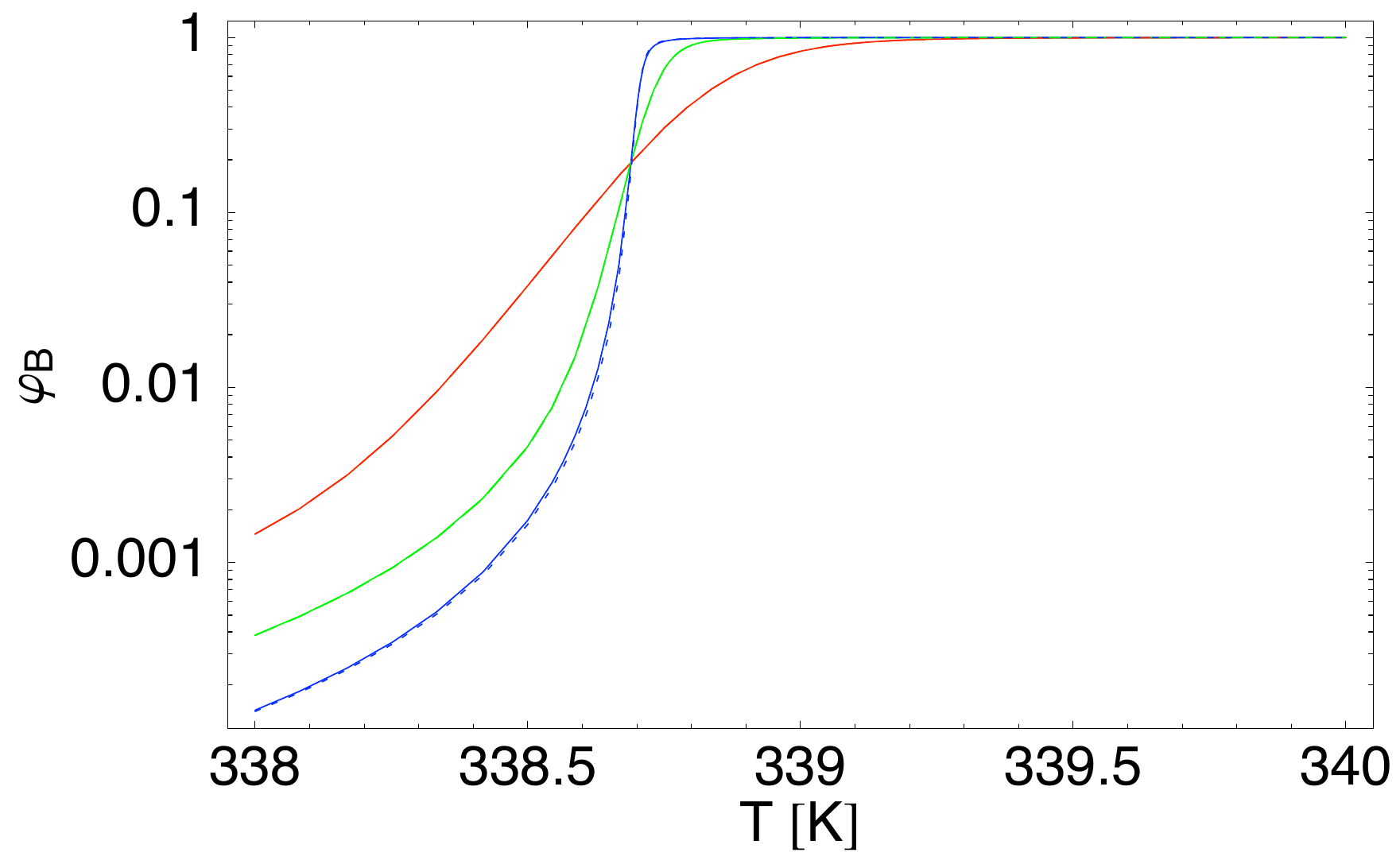}(d)
\caption{Melting curves: comparison of the exact result with the
one-sequence approximation (including completely open state) for a
free chain (no loop entropy, no sliding). Exact results (solid
curves) from right to left near the upper part of the curves
($T>339$~K), $N = 500$ , 2000, 10000; one-sequence
approximation,  $N = 500$ (long dashed curve), 2000 (intermediate
dashed curve), 10000 (short dashed curve), $f=0.5$ and other parameters
as in Figure~\ref{fig3}). (a) $J=4.57$~kJ/mol; (b) as in (a) but now
Linear-Log plot; (c) $J=9.13$~kJ/mol; (d) as in (c) but now
Linear-Log plot. In (c) and (d), the dashed and solid curves are superposed.} \label{fig10}
\end{center}
\end{figure}

The above one-sequence approximation should be valid for
sufficiently short chains. After determining its range of validity
when loop entropy is neglected, we can then use it with confidence
within this range to examine the influence of loop entropy on DNA
denaturation.
In Fig.~\ref{fig10} we test the validity of the one-sequence approximation
with neither loop entropy, nor chain sliding by comparing it with
exact result (\ref{phin}) for which the partition function includes the
completely open state. From now on we fix the weighting factor $f$ at 0.5, which, as explained earlier, is close to the one estimated from experiment.  We observe that the one sequence approximation is accurate when $N\leq500$ for $J=4.57$~kJ/mol (Fig.~\ref{fig9}a)  and accurate beyond $N\leq10000$ for $J=9.13$~kJ/mol (Fig.~\ref{fig9}b); in both cases
studied the melting temperature is well reproduced, although the
transition width is underestimated for $J=4.57$~kJ/mol when  $N\leq500$ (with the
discrepancy increasing with increasing $N$). The one-sequence
approximation also somewhat overestimates the temperature $T^*$ at
which the melting curves intersect. We conclude that the limiting value of $N$ for which
the one-sequence approximation is accurate depends critically on the value of $J_0$ via the loop entropy factor(\ref{lif}).

\begin{figure}[t]
\begin{center}
\includegraphics[height=6cm]{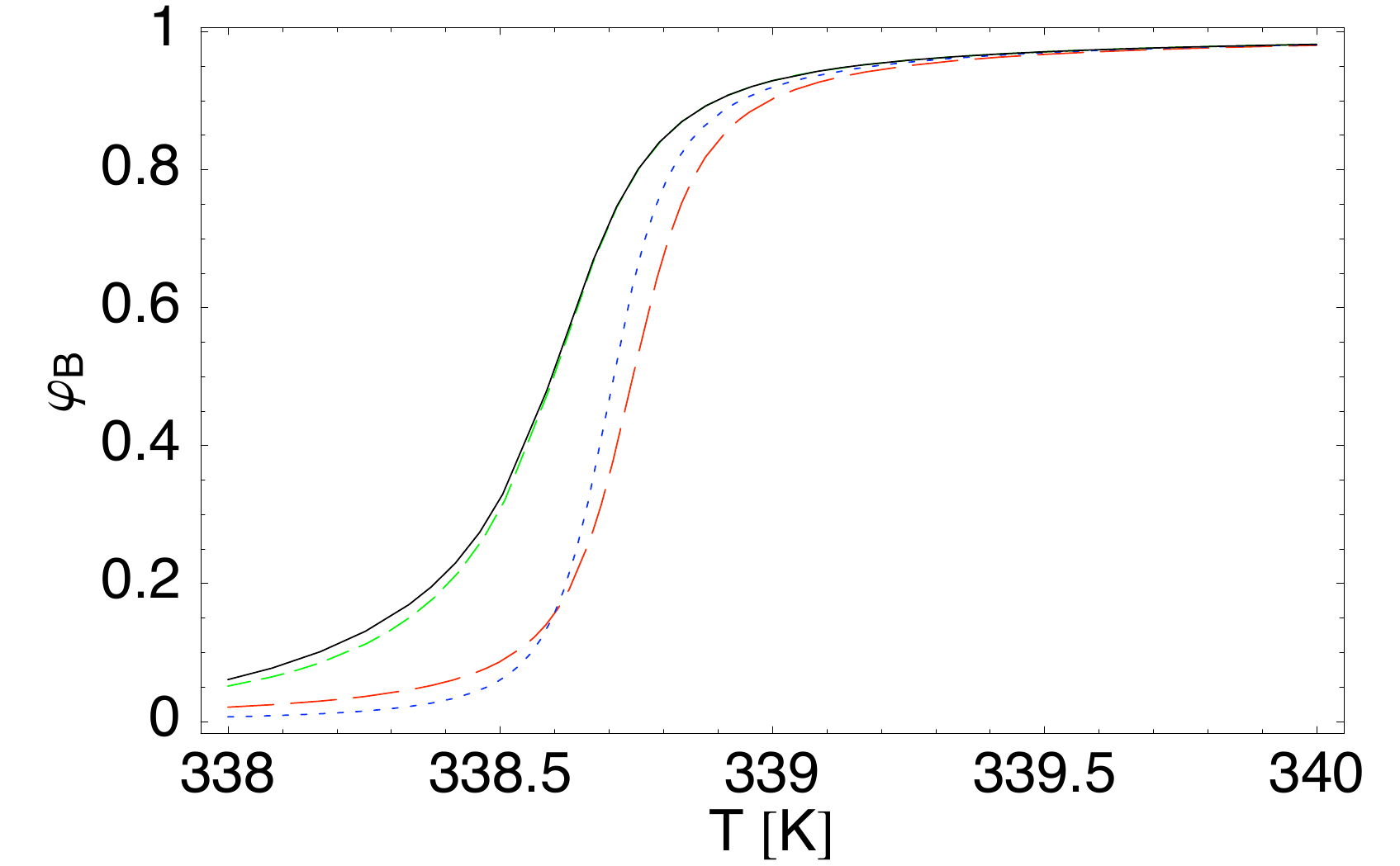}(a)\\
\includegraphics[height=6cm]{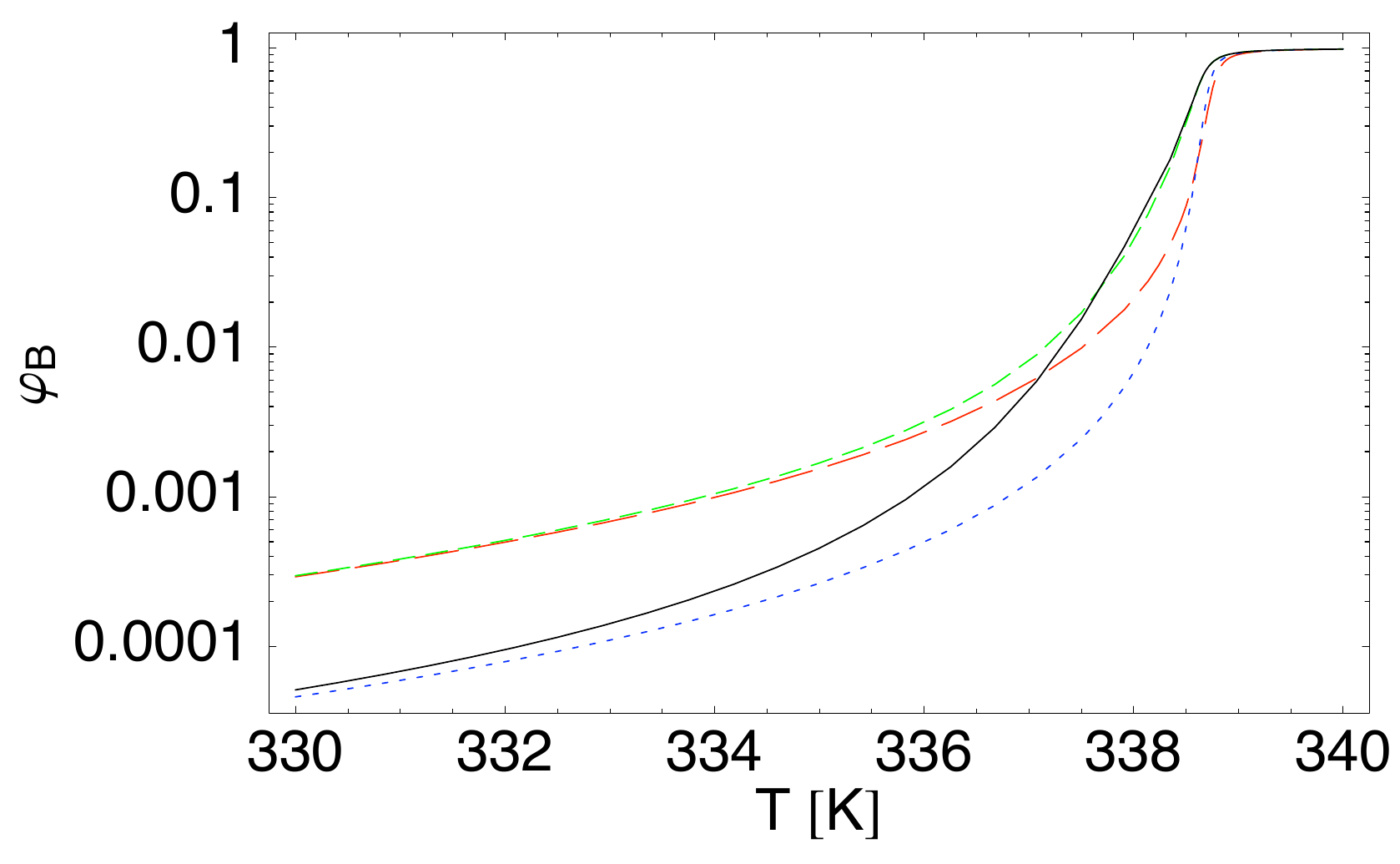}(b)
\caption{Internal melting curves (associated chains): Comparison of
various one-sequence approximations for a free chain with $N=2000$:
(a) Linear plot (b) Linear-Log plot; neither loop entropy  nor
sliding (long dashed curve), sliding only (intermediate dashed
curve), loop entropy only (short dashed curve), both loop entropy
and sliding (solid curve), ($k = 1.7$, $n_0 = 198$, $f=0.5$, other parameters as
in Figure~\ref{fig3}).} \label{fig11}
\end{center}
\end{figure}

Because we are now interested in studying the effects of loop
entropy on thermal denaturation, we employ the smaller value for $J$
(4.57~kJ/mol). Despite this smaller value, the inclusion of loop entropy reinforces the validity of the one-sequence approximation. For $J=4.57$~kJ/mol, $k=1.7$, and $D=100$ ($n_0=198$),  the readjusted value $\hat{J}$ (\ref{jp}) is greater than  9.13~kJ/mol, implying that in this case bubbles are even more highly suppressed for $J=4.57$~kJ/mol with loop entropy than for $J=9.13$~kJ/mol without loop entropy. In Fig.~\ref{fig11} we observe that at low temperature the chain-sliding-only model gives the
highest melting and the loop-entropy one the lowest. At higher
temperature the sliding-loop entropy model gives the highest
melting. For the case  considered in Fig.~\ref{fig11}, we therefore expect the accuracy of the one-sequence approximation to be comparable to that seen in Fig.~\ref{fig10}c,d (and not Fig.~\ref{fig10}a,b).

\begin{figure}[ht]
\begin{center}
\includegraphics[height=6cm]{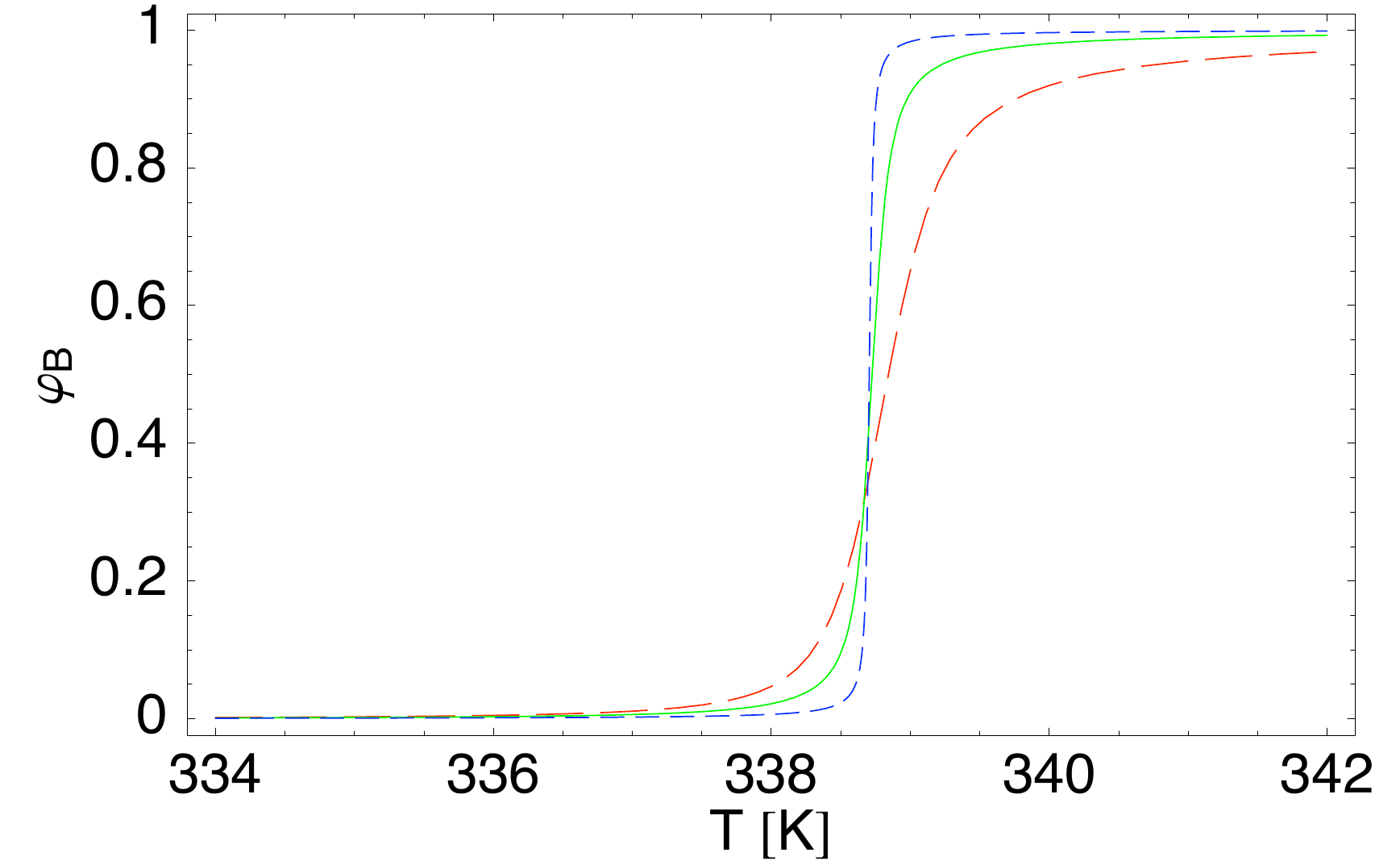}(a)\\
\includegraphics[height=6cm]{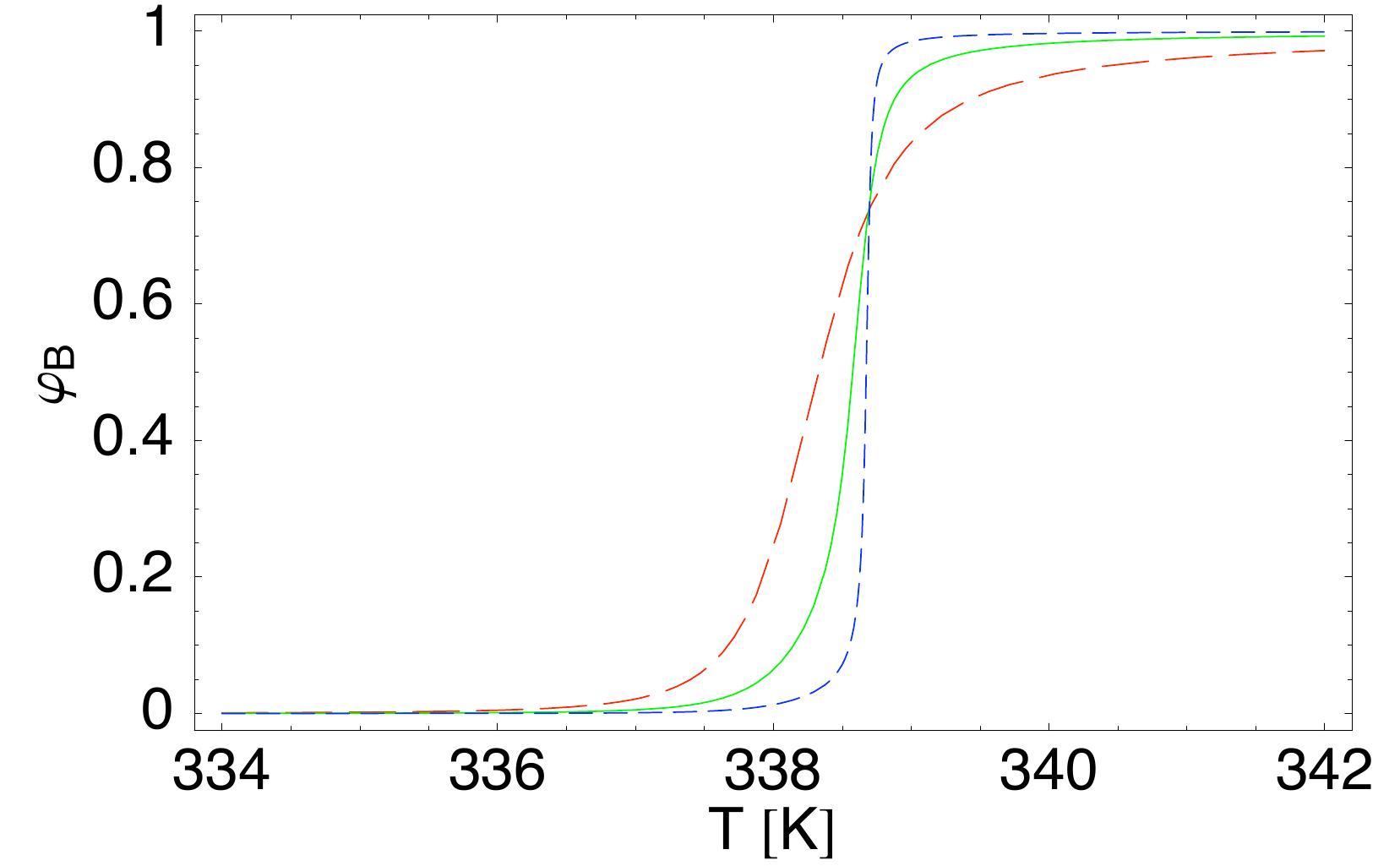}(b)
 \caption{Internal melting curves (associated chains) obtained
using the Loop Entropy-Sliding model for free chains of three
different lengths: $N = 500$ (long dashed curve); 2000 (solid
curve); 10000 (short dashed curve) with $J=4.57$~kJ/mol, $k = 1.7$, $n_0
= 198$, $f=0.5$ (other parameters as in Figure~\ref{fig3}): (a) with
neither loop entropy, nor chain sliding; (b) with  loop entropy and
chain sliding.} \label{fig12}
\end{center}
\end{figure}

In Fig.~\ref{fig12} we plot the melting curves using the \emph{Loop Entropy-Sliding}
model for free chains of three different lengths ($N = 500, 200,
10000$) and compare the results obtained without loop entropy and
sliding. We note that due to the combined effects of sliding and
loop entropy the melting temperature increases with increasing $N$
and the width of the transition decreases (Fig.~\ref{fig12}b), in agreement
with experiment \cite{blake} (for $f=0.5$ the temperature $T^*$ at which the
melting curves intersect is now greater than $T_m(N)$, the opposite
of what occurs when loop entropy and sliding are neglected,
see Fig.~\ref{fig12}a). The model prediction for the difference between the
melting temperatures for $N=500$ and 10000 is about 0.5~K (the results for $N>10000$
should be very close to  the  $N=10000$ one).
When chain dissociation is added to the model, one can reasonably
expect that the melting temperature for $N=500$ will decrease by
about 0.5~K \cite{PS, PS1} and that for $N\geq10000$ will hardly
change. This result suggests that once chain dissociation is
incorporated into the current model, it should be possible to
account for the experimental results of \cite{blake}
[$T_m(30000)-T_m(500)\simeq 1 $~K and decreasing transition width as
$N$ increases].


\section{Concluding remarks}


This paper presents the extension of a theoretical model of DNA
denaturation~\cite{JMN,JMN2} that couples the base-pair states,
unbroken or broken, and the chain configurational degrees of
freedom. The elastic contributions are taken into account, arising
from chain bending, torsional and stretching rigidities, the values
of which depend on the neighboring base-pair states. The difference
of bond lengths in ssDNA (0.34 nm) and dsDNA (0.71 nm) is also
included in the Hamiltonian. This model, tackled by analytical
means, provides new insight into the dependence of the effective Ising
parameters, used in previous Ising-like models, on microscopic
elastic moduli. The main conclusion is that all these features lead
to a renormalization of the bare Ising parameters on the order of
magnitude of the thermal energy. Hence, they cannot be ignored when
relating microscopic properties, extracted for example from {\it ab
initio} calculations or experiments on DNA fragments, to the
collective properties of the whole chain measured, for instance, in
single DNA molecule experiments (atomic force microscopy, optical
and magnetic tweezers, tethered particle motion). As an
illustration, without considering the effects of stretching
elasticity and base-pair length, the energy cost to open a
base-pair, $2\mu$, would be directly related to the same quantity
measured with a force apparatus~\cite{pincet,footnote4}. But
$\mu$ is renormalized by these effects and is lowered by 0.5 to 1
$k_BT$ when the bare value is close to 2 $k_BT$. The same conclusion
holds for the destacking, $J$,  or stacking, $K$, parameters.

In this work, we also analyze finite size effects. In particular the
role of closed boundary conditions on melting curves for finite
lengths is investigated in order to model a clamped polydA-polydT
DNA inserts. Two approximations are considered: (i) the one-sequence
approximation amounts to neglecting  configurations with several bubbles
and (ii) the two-state one keeps only the contributions from the
completely closed and open chains~\cite{blake}. In the range of
parameters studied, the agreement with the exact result is excellent
in case (i), whereas it is much less satisfactory in case (ii).
We also undertake the integration of loop entropy in case (i), which  leads to an
increase in $T_m$ that is associated with
the loop entropy cost and depends on the value of the loop entropy chain stiffness parameter $D$ (for $N\sim100$ there is a shift of 1~K for $D=100$ and of 5~K for $D=1$).  Finally, we study free polymers chains using exact results with neither loop entropy nor chain sliding and the one-sequence approximation with  loop entropy and chain sliding. Our major conclusion is that the experimentally  observed increase in $T_m$ with increasing chain length for homopolymers can be accounted for by incorporating both loop entropy and chain sliding into our model.  The simplicity of our method of incorporating  loop entropy  into the one-sequence approximation paves the way to a deeper study of the role of chain stiffness in the loop entropy factor, $g_{\rm LE}$. We underline that careful experiments on free and clamped homopolymers of different lengths (in solution or in single molecule experiments) would be extremely useful in elucidating the role of DNA finite size effects. 

From an experimental perspective, our findings are relevant for free
DNA in dilute solutions, without any constraint on chain
configurations, nor any applied force or torque. An ingredient that
we did not consider so far is the gain in translational entropy due
to strand separation in the case of dissociation~\cite{wartell}.
A correct treatment of this mechanism  consists in writing a chemical
equilibrium between completely denatured single strands and
partially bound ones (work  in progress).

The case of constrained DNA is more involved. If a force or a torque
is applied, for instance in tweezer experiments, rotational symmetry
is lost in the Hamiltonian, which prevents an analytical solution of
the problem.  Numerical or approximate schemes, such as variational
principles, may be used. Another interesting constraint concerns
polymer looping~\cite{benham}. Circular DNAs appear in the case of
transposons or insertion sequences~\cite{cell_book,pouget2}. Writing
down the polymer closure (e.g., for the determination of the $J$-factor)
is a formidable task because it corresponds to the global constraint
$\sum {\mathbf t}_i=0$, formally equivalent to an applied
force~\cite{marko}. We can, however,  partially take into account
looping in our framework by imposing periodic boundary conditions on
the vectors $\hat{\mathbf e}_{\mu,i}$ and/or on $\sigma_i$, instead
of the end condition $| V \rangle$. This can be handled using  the
transfer matrix  method. In the case of superhelical twist, the
polymer winds one or several times around its tangent vectors
${\mathbf t}_i$. This condition can also be enforced \textit{via}  the  boundary
conditions, by requiring that the appropriate combination of Euler
angles acquires a phase multiple of $2\pi$ when going from $i=N$ to
$i=1$. This topological constraint should lead to an increased
fraction of denaturated base-pairs, in order to release the
torsional energy cost, and consequently to an increased flexibility,
thereby facilitating cyclization. Our predictions for the end-to-end
distance can also be compared to experiment, because $R$ is
proportional to the radius of gyration, which can be measured in viscosity
experiments.

All the results presented in this paper concern homopolynucleotides
and the numerical applications focused on PolydA-dT. This
work can, however, be generalized to heteropolymers, although  a minimal amount of numerical
work is necessary to handle the reduction of the transfer matrices. Nonetheless, a numerical study of heteropolymers
would require the knowledge of the microscopic elastic moduli, which
are far from being known with any certainty for any pair of the four nucleotides A, T,
G and C.

\section*{Appendix}

In this Appendix we extract smooth melting curves from the
experimental data in~\cite{blake}. For the poly dA-dT DNA polymer
with free ends and  30000 base-pairs we have used the temperature
derivative of
\begin{equation} \label{phifitinf}
\varphi_{\rm fit} = \frac{c_f}{2} \left[ 1 - \frac{\sinh(-a_f+\beta
\mu_f)}{\sqrt{e^{-4\beta J_f}+\sinh^2(-a_f+\beta \mu_f)}} \right],
\end{equation}
where $c_f$, $a_f$, and $\mu_f$ are fitting parameters (simplified
$N=\infty$ Ising form); this  functional form arises in simple Ising
models of DNA denaturation~\cite{montroll}.

\begin{figure}[ht]
\begin{center}
\includegraphics[height=6cm]{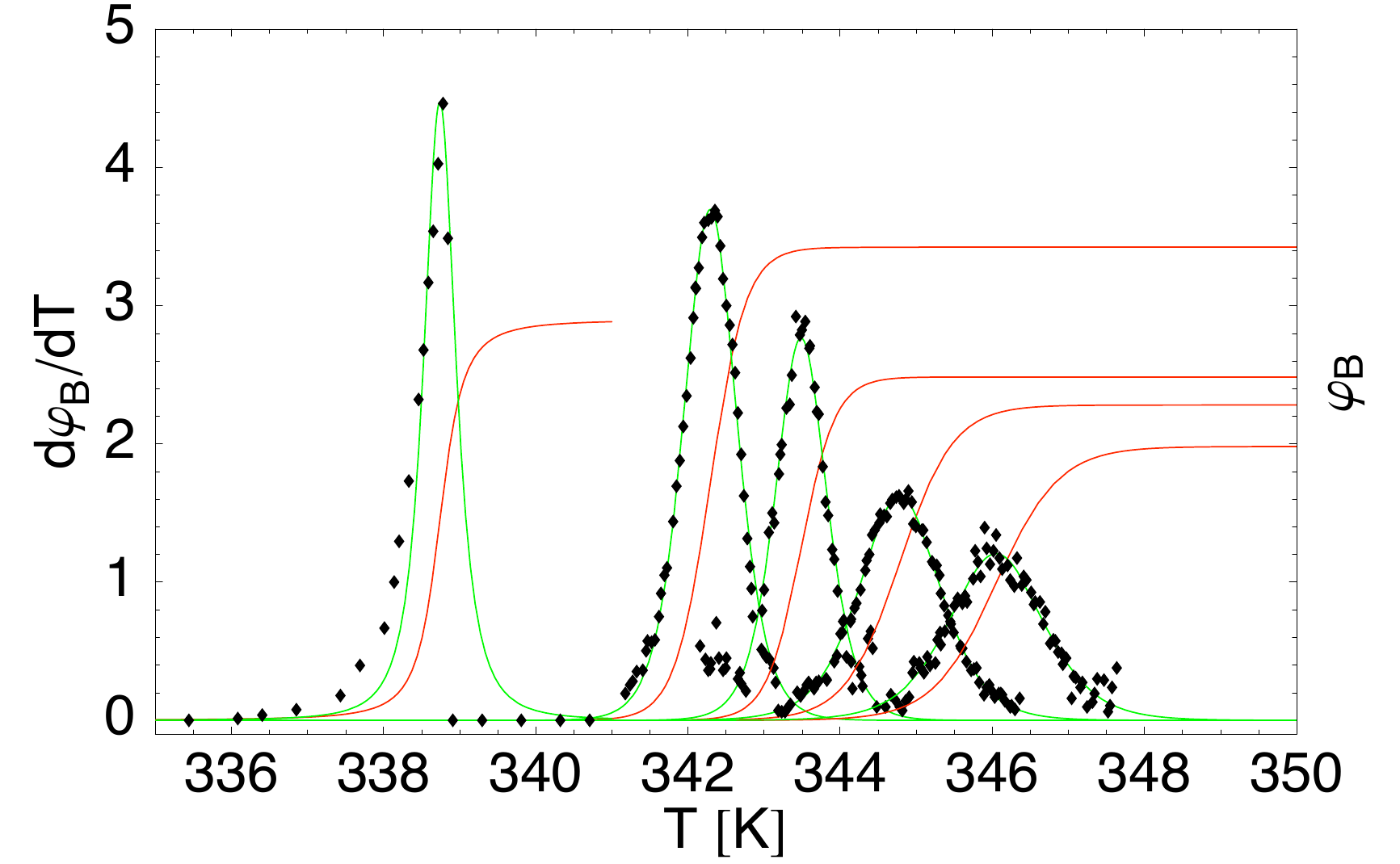}
\caption{Absorbance temperature derivative (un-normalized
$d\varphi_B/dT$) vs. temperature: experimental data
points~\cite{blake} and un-normalized fitted functions (green
curves, left y-axis); UV absorbance (un-normalized fraction of
broken base-pairs, $\varphi_B$), vs. temperature (red curves, right
y-axis) (from left to right: $N=30000$, 136, 105, 83, and 67).}
\label{figblakedata}
\end{center}
\end{figure}

\begin{figure}[ht]
\begin{center}
\includegraphics[height=6cm]{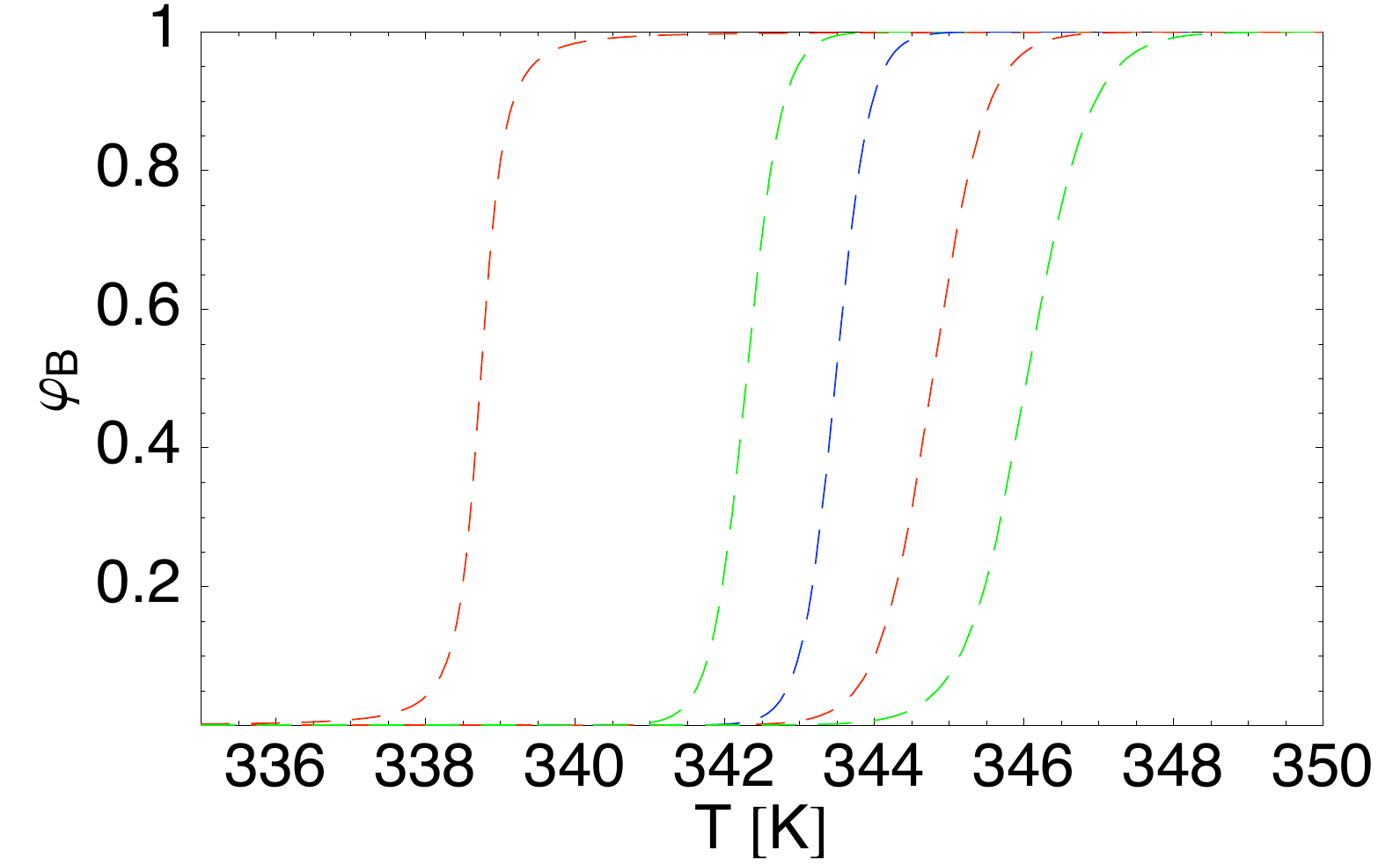}(a)
\includegraphics[height=6cm]{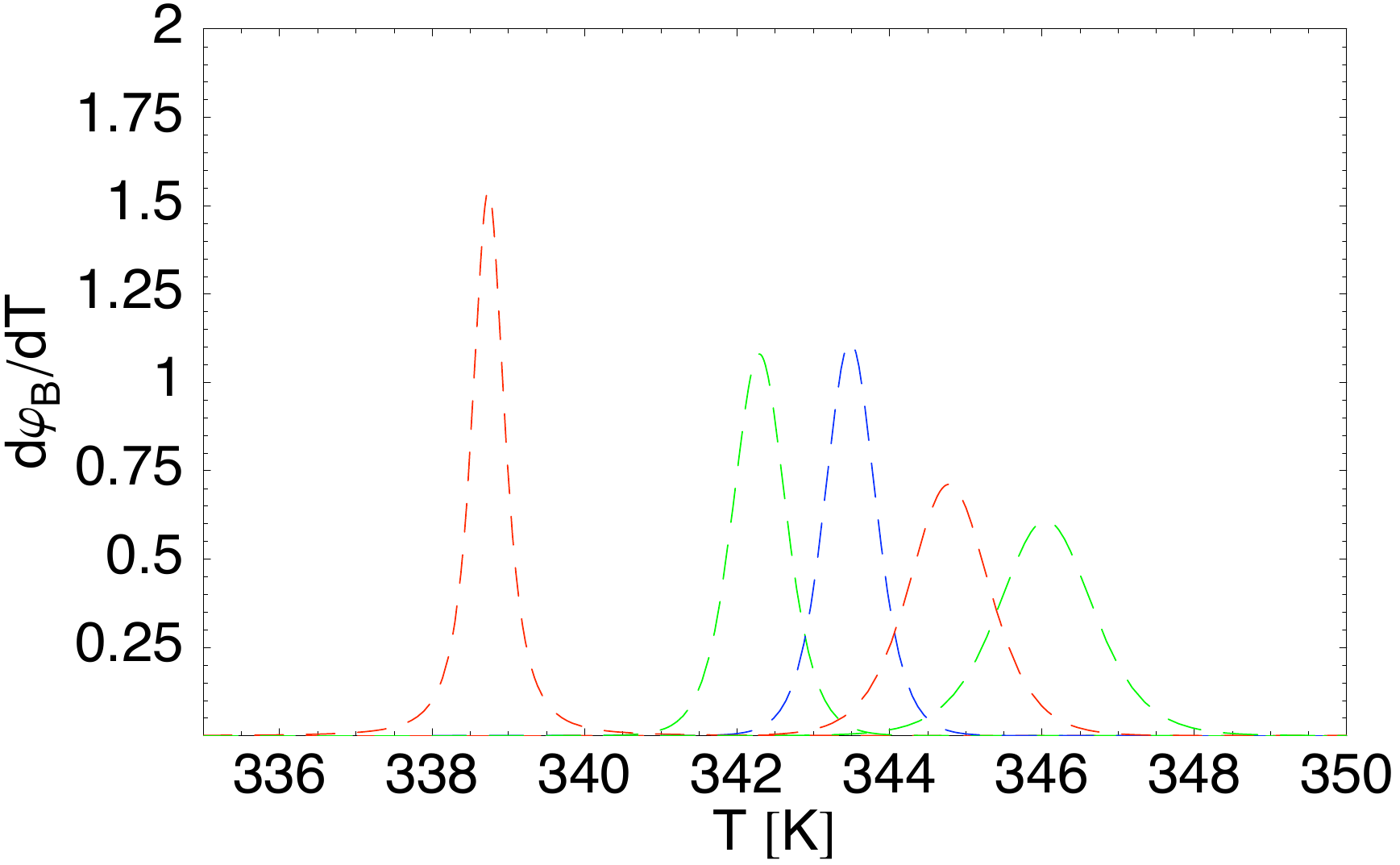}(b)
\caption{Normalized functions fitted to the experimental data~\cite{blake}: (a) fraction of broken base-pairs vs. temperature, $\varphi_B$; (b) $d\varphi_B/dT$ vs. temperature (from left to right, $N=30000$, 136, 105, 83, and 67).} 
\label{figblakefitdata}
\end{center}
\end{figure}

For A-T inserts we have used the temperature derivative of
\begin{equation} \label{phifitinsert}
\varphi_{\rm fit} = \frac{c_f}{2 h_f} \left\{ 1 - \tanh[h_f(T_f -
T)] \right\},
\end{equation}
where $c_f$, $h_f$, and $T_f$ are fitting parameters; this
functional form arises in a two-state treatment of simple Ising
models of DNA denaturation~\cite{PS} (the use of the two-state form
here to extract smooth experimental melting curves does not imply
that the two-state approximation is a valid one, see Fig.~\ref{fig6}). As
shown in Fig.~\ref{figblakedata} the areas under the fitted $d\varphi_B/dT$ functions
are not normalized to one. We thus assume that the normalized fitted
$\varphi_B$ functions (Fig.~\ref{figblakefitdata}) represent a good approximation to the
fraction of open base-pairs for the A-T segments. By examining
Fig.~A.1 we see that this assumption is well borne out for the A-T
inserts, but less so for the $N = 30000$ base-pair chain because of
difficulties in reading the data off the experimental curve and the
asymmetry of this curve. Our choice of fitting functions give
symmetric curves about the melting temperature and thus cannot
accounted for the observed asymmetry for $N = 30000$. The observed
asymmetry probably cannot be explained by loop entropy  and chain sliding (for infinite chains at least) because when they are  included in the model, the melting curves becomes flatter to the left of the melting temperature and steeper to the right, the
opposite of what is observed in Fig.~\ref{figblakedata} (for finite chains, however, the combined effects of loop entropy and chain sliding can be different, see Fig.~\ref{fig11}). Although the $N = 30000$ base-pair chain melting temperature $\sim 339$~K is well reproduced, the width of the transition appears to be overestimated. The general trend is for both the melting temperature and transition width to decrease with increasing $N$. As the length of the insert increases the melting should tend to the infinite free chain result.

\begin{thebibliography}{99}
\bibitem{footnote1}
The rotation per base-pair (or equilibrium twist) being $\Phi_0=0.19\,\pi$~rad~\cite{peck} (the pitch is then $p=3.55$~nm), and the radius $r=1.19$~nm, one can estimate the monomer size in the fully extended ssDNA form as $a_{ss}=\Phi_0\sqrt{r^2+(p/2\pi)^2}\approx0.79$~nm which is larger than $a_{ds}$. It should be noticed however, that ssDNA is usually not fully extended and has an helical shape too.
\bibitem{footnote3}
This result does not change if we include in the model the equilibrium twist $\Phi_0=0.19\,\pi$~rad for dsDNA in the Hamiltonian~(\ref{DHWC}) by changing $\cos(\phi+\psi)$ in $\cos(\phi+\psi-\Phi_0)$ for unbroken nearest-neighbours.
\bibitem{footnote4}
Even though the (screened) electrostatic repulsion of the phosphate atoms of the
sugar DNA backbone is not taken into account in those experiments and is also likely to decrease the real $\mu$~\cite{perez}.
\end{thebibliography}
\end{document}